\documentclass[aps,prl,twocolumn,,superscriptaddress,floatfix]{revtex4-1}
\usepackage[svgnames]{xcolor}
\usepackage[colorlinks=true,urlcolor = purple,linkcolor=teal,citecolor=magenta,bookmarks=false]{hyperref}
\usepackage{amsfonts,amsmath,amssymb}
\usepackage{graphicx}
\usepackage{dsfont}
\usepackage{graphicx}
\usepackage{amssymb}
\usepackage{amsmath}
\usepackage{empheq}
\usepackage{bm}
\usepackage{ulem}
\usepackage{stmaryrd}
\usepackage{tkz-graph}
\usepackage{tikz}
\usetikzlibrary{arrows,%
                petri,%
                topaths}%
\usepackage{tkz-berge}
\usetikzlibrary{matrix,shapes,arrows,decorations.markings}
\usepackage[nomessages]{fp}
\newcommand{\bc}{\begin{center}}
\newcommand{\ec}{\end{center}}
\newcommand{\beq}{\begin{equation}}
\newcommand{\eeq}{\end{equation}}
\newcommand{\beqa}{\begin{eqnarray}}
\newcommand{\eeqa}{\end{eqnarray}}
\newcommand{\beqs}{\begin{eqnarray*}}
\newcommand{\eeqs}{\end{eqnarray*}}
\newcommand{\bi}{\begin{itemize}}
\newcommand{\ei}{\end{itemize}}

\def\vev#1{\langle #1 \rangle}

\newcommand{\Tr}{{\rm Tr}}

\tikzset{
    partial ellipse/.style args={#1:#2:#3}{
        insert path={+ (#1:#3) arc (#1:#2:#3)}
    }
}

\tikzset{middlearrow/.style={
        decoration={markings,
            mark= at position 0.55 with {\arrow[thick]{#1}} ,
        },
        postaction={decorate}
    }
}

\newcommand{\singleloop}[3] {
\begin{tikzpicture}[scale=#3,baseline={([yshift=-3pt]current bounding box.center)}]
\draw[middlearrow={to}] (-1.3,0) arc (180:90:1.3cm and 0.8cm);
\draw[middlearrow={to}] (0,0.8) arc (90:0:1.3cm and 0.8cm);
\draw[middlearrow={to}] (1.3,0) arc (0:-90:1.3cm and 0.8cm);
\draw[middlearrow={to}] (0,-0.8) arc (-90:-180:1.3cm and 0.8cm);
\draw [fill=gray!30] (-1.3,0) circle (0.3);
\node[scale=#3] at (-1.3,0) {A};
\draw [fill=gray!30] (1.3,0) circle (0.3);
\node[scale=#3] at (1.3,0) {B};
\draw [fill] (0,0.8) circle (0.1);
\node[below,scale=1.3*#3] at (0,0.75) {$#1$};
\draw [fill] (0,-0.8) circle (0.1);
\node[above,scale=1.3*#3] at (0,-0.75) {$#2$};
\end{tikzpicture}}

\newcommand{\contact}[2]{
\begin{tikzpicture}[scale=#2,baseline={([yshift=-3pt]current bounding box.center)}]
    \draw[middlearrow={to}] (-1,0) arc (180:0:1cm and 0.8cm);
    \draw[middlearrow={to}] (1,0) arc (0:-180:1cm and 0.8cm);
    \draw [fill=gray!30] (-1,0) circle (0.3);
    \node[scale=#2] at (-1,0) {A};
    \draw[middlearrow={to}] (1,0) arc (180:0:1cm and 0.8cm);
    \draw[middlearrow={to}] (3,0) arc (0:-180:1cm and 0.8cm);    
    \draw [fill=gray!30] (3,0) circle (0.3);
    \node[scale=#2] at (3,0) {B};
    \draw [fill] (1,0) circle (0.1); 
    \node[right,scale=1.5*#2] at(1.1,0) {$#1$};
\end{tikzpicture}}

\newcommand{\separes}[3]{
\begin{tikzpicture}[scale=#3,baseline={([yshift=-3pt]current bounding box.center)}]
    \draw[middlearrow={to}] (-1,0) arc (180:0:1cm and 0.8cm);
    \draw[middlearrow={to}] (1,0) arc (0:-180:1cm and 0.8cm);    
    \draw [fill=gray!30] (-1,0) circle (0.3);
    \node[scale=#3] at (-1,0) {A};
    \draw[middlearrow={to}] (1.3,0) arc (180:0:1cm and 0.8cm);
    \draw[middlearrow={to}] (3.3,0) arc (0:-180:1cm and 0.8cm);    
    \draw [fill=gray!30] (3.3,0) circle (0.3);
    \node[scale=#3] at (3.3,0) {B};
    \draw [fill] (1.3,0) circle (0.1) node[right,scale=1.5*#3] {$#2$};
    \draw [fill] (1,0) circle (0.1) node[left,scale=1.5*#3] {$#1$};
\end{tikzpicture}} 

\newcommand{\oneloopg}[2]{
\begin{tikzpicture}[scale=#2,baseline={([yshift=-3pt]current bounding box.center)}]
    \draw[middlearrow={to}] (-1,0) arc (180:0:1cm and 0.8cm);
    \draw[middlearrow={to}] (1,0) arc (0:-180:1cm and 0.8cm);    
    \draw [fill=gray!30] (-1,0) circle (0.3);
    \node[scale=#2] at (-1,0) {A};
    \draw [fill] (1,0) circle (0.1) node[right,scale=1.5*#2] {$#1$};
\end{tikzpicture}} 

\newcommand{\oneloopd}[2]{
\begin{tikzpicture}[scale=#2,baseline={([yshift=-3pt]current bounding box.center)}]
    \draw[middlearrow={to}] (0,0) arc (180:0:1cm and 0.8cm);
    \draw[middlearrow={to}] (2,0) arc (0:-180:1cm and 0.8cm);    
    \draw [fill=gray!30] (2,0) circle (0.3);
    \node[scale=#2] at (2,0) {B};
    \draw [fill] (0,0) circle (0.1) node[left,scale=1.5*#2] {$#1$};
\end{tikzpicture}} 

\newcommand{\graphN}[2]
{
\FPeval{\a}{round(#2 *(-.23),2)}
\FPeval{\b}{round(#2 *(-.1),2)}
\FPeval{\c}{round(#2 * (.22),2)}
\FPeval{\d}{round(#2 *(0.67),2)}
\FPeval{\e}{round(#2 *(2)-#2*#2/2 ,2)}
\begin{tikzpicture}[baseline={([yshift=-3pt]current bounding box.center)}]
\SetVertexSimple
\useasboundingbox (\a,\b) rectangle (\c,\d);
\tikzset{VertexStyle/.append style={minimum size=2pt, inner sep=1pt}}
\Vertex{a} 
\node[left,scale=1.7*#2] at (0,0) {$#1$};
\begin{scope}[decoration={markings,mark = at position 0.55 with {\arrow[scale=\e]{to}}}]
\Loop[dist=#2 cm,dir=NO,style={thin,postaction={decorate}}](a)
\end{scope}
\end{tikzpicture}
}

\newcommand{\graphNcroixdeux}[3]
{
\graphN{#1}{#3} \hspace{3.5pt} \graphN{#2}{#3}
}

\newcommand{\graphNcroixtrois}[4]
{
\graphN{#1}{#4} \hspace{3.5pt} \graphN{#2}{#4} \hspace{3.5pt} \graphN{#3}{#4}
}

\newcommand{\graphNdeux}[2]
{
\FPeval{\a}{round(#2 *(-.65),2)}
\FPeval{\b}{round(#2 *(-.265),3)}
\FPeval{\c}{round(#2 * (.66),2)}
\FPeval{\d}{round(#2 *(.24),2)}
\FPeval{\e}{round(#2 *(2)-#2*#2/2 ,2)}
\begin{tikzpicture}[baseline={([yshift=-3pt]current bounding box.center)}]
\GraphInit[vstyle=Classic]
\SetVertexSimple
\useasboundingbox (\a,\b) rectangle (\c,\d); 
\tikzset{VertexStyle/.append style={minimum size=2pt, inner sep=1pt}}
\Vertex{a}
\node[below,scale=1.7*#2] at (0,0) {$#1$};
\begin{scope}[decoration={markings,mark = at position 0.55 with {\arrow[thin, scale=\e]{to}}}]
\Loop[dist=#2 cm,dir=WE,style={thin,postaction={decorate}}](a)
\Loop[dist=#2 cm,dir=EA,style={thin,postaction={decorate}}](a)
\end{scope}
\end{tikzpicture}
}

\newcommand{\graphFdeux}[3]
{
\begin{tikzpicture}[baseline={([yshift=0pt]current bounding box.center)},scale=#3]
\GraphInit[vstyle=Classic]
\SetVertexSimple
\tikzset{VertexStyle/.append style={minimum size=2pt, inner sep=1pt}}
\Vertex{a} \node[below,scale=1.7*#3] at (0,0) {$#1$};
\EA(a){b} \node[below,scale=1.7*#3] at(1,0) {$#2$};
\begin{scope}[decoration={markings,mark = at position 0.55 with {\arrow[thin, scale=#3*2]{to}}}]
\tikzset{EdgeStyle/.style = {thin,postaction={decorate},bend left=75}}
\Edge(b)(a)
\Edge(a)(b)
\end{scope}
\end{tikzpicture}
}
\newcommand{\graphFdeuxpointN}[3]
{
\begin{tikzpicture}[baseline={([yshift=0pt]current bounding box.center)},scale=#3]
\GraphInit[vstyle=Classic]
\SetGraphUnit{#3}
\SetVertexSimple
\tikzset{VertexStyle/.append style={minimum size=2pt, inner sep=1pt}}
\Vertex{a}
\node[below,scale=#3*1.2]{$#1$};
\EA(a){b}
\node[below,scale=#3*1.2] at (#3*1.1,0){$#2$};
\begin{scope}[decoration={markings,mark = at position 0.55 with {\arrow[thin, scale=#3*2]{to}}}]
\tikzset{EdgeStyle/.style = {thin,postaction={decorate},bend left=75}}
\Edge(b)(a)
\Edge(a)(b)
\tikzset{EdgeStyle/.style = {thin,postaction={decorate}}}
\Loop[dist=#3 cm,dir=EA,style={thin,postaction={decorate}}](b)
\end{scope}
\end{tikzpicture}
}

\newcommand{\graphNpointFdeux}[3]
{
\begin{tikzpicture}[baseline={([yshift=0pt]current bounding box.center)},scale=#3]
\GraphInit[vstyle=Classic]
\SetGraphUnit{#3}
\SetVertexSimple
\tikzset{VertexStyle/.append style={minimum size=2pt, inner sep=1pt}}
\Vertex{a}
\node[below,scale=#3*1.2]{$#1$};
\EA(a){b}
\node[below,scale=#3*1.2] at (#3*1.1,0){$#2$};
\begin{scope}[decoration={markings,mark = at position 0.55 with {\arrow[thin, scale=#3*2]{to}}}]
\tikzset{EdgeStyle/.style = {thin,postaction={decorate},bend left=75}}
\Edge(b)(a)
\Edge(a)(b)
\tikzset{EdgeStyle/.style = {thin,postaction={decorate}}}
\Loop[dist=#3 cm,dir=WE,style={thin,postaction={decorate}}](a)
\end{scope}
\end{tikzpicture}
}
\newcommand{\graphNdeuxcroixN}[3]
{
\graphNdeux{#1}{#3} \hspace{3.5pt} \graphN{#2}{#3}
}

\newcommand{\graphNcroixNdeux}[3]
{
\graphN{#1}{#3} \hspace{3.5pt} \graphNdeux{#2}{#3}
}

\newcommand{\graphNcroixFdeux}[4]
{
\graphN{#1}{#4} \hspace{3.5pt} \graphFdeux{#2}{#3}{#4}
}

\newcommand{\graphNcroixmFdeux}[4]
{
\begin{tikzpicture}[baseline={([yshift=-3pt]current bounding box.center)},scale=#4]
\GraphInit[vstyle=Classic] 
\SetVertexSimple
\tikzset{VertexStyle/.append style={minimum size=2pt, inner sep=1pt}}
\Vertex{a}
\node[below,scale=#4*1.5]{$#1$};
\EA(a){b}
\node[below,scale=#4*1.5] at (1.1,0){$#3$};
\Vertex[x=0.5,y=0]{c}
\node[below,scale=#4*1.5] at (0.7,-0.2){$#2$};
\begin{scope}[decoration={markings,mark = at position 0.55 with {\arrow[thin, scale=#4*1.7]{to}}}]
\tikzset{EdgeStyle/.style = {thin,postaction={decorate},bend left=75}}
\Edge(b)(a)
\Edge(a)(b)
\tikzset{EdgeStyle/.style = {thin,postaction={decorate}}}
\Loop[dist=1cm,dir=NO,style={thin,postaction={decorate}}](c)
\end{scope}
\end{tikzpicture}
}

\newcommand{\graphFdeuxcroixN}[4]
{
 \graphFdeux{#1}{#2}{#4} \hspace{7pt} \graphN{#3}{#4}
}
\newcommand{\graphNtrois}[2]
{
\begin{tikzpicture}[baseline={([yshift=0pt]current bounding box.center)},scale=#2]
\GraphInit[vstyle=Classic]
\SetVertexSimple
\tikzset{VertexStyle/.append style={minimum size=2pt, inner sep=1pt}}
\Vertex{a}
\node[below,scale=#2*1.7]{$#1$};
\begin{scope}[decoration={markings,mark = at position 0.57 with {\arrow[thin, scale=#2*2.3]{to}}}]
\Loop[dist=1 cm,dir=NO,style={thin,postaction={decorate}}](a)
\Loop[dist=1 cm,dir=EA,style={thin,postaction={decorate}}](a)
\Loop[dist=1 cm,dir=WE,style={thin,postaction={decorate}}](a)\end{scope}
\end{tikzpicture}
}
\newcommand{\graphFtrois}[4]
{
\begin{tikzpicture}[baseline={([yshift=-3pt]current bounding box.center)},scale=#4]
\GraphInit[vstyle=Classic]
\SetVertexSimple
\tikzset{VertexStyle/.append style={minimum size=2pt, inner sep=1pt}}
\begin{scope}[decoration={markings,mark = at position 0.55 with {\arrow[thin, scale=#4*4]{to}}}]
\tikzset{EdgeStyle/.style = {thin,postaction={decorate},bend right}}
\Vertices{circle}{$#3$,$#2$,$#1$}
\node[right,scale=#4*3] at(1,0){$#2$};
\node[left,scale=#4*3] at(-0.9,1){$#3$};
\node[left,scale=#4*3] at(-0.9,-1){$#1$};
\Edges($#3$,$#2$,$#1$,$#3$)
\end{scope}
\end{tikzpicture}
}

\newcommand{\graphFquatre}[5]
{ 
\begin{tikzpicture}[baseline={([yshift=0pt]current bounding box.center)},scale=#5]
\SetVertexSimple
\tikzset{VertexStyle/.append style={minimum size=2pt, inner sep=1pt}}
\begin{scope}[decoration={markings,mark = at position 0.55 with {\arrow[thin, scale=#5*2.5]{to}}}]
\tikzset{EdgeStyle/.style = {thin,postaction={decorate},bend right}}
\Vertex[x=-0.7,y=-0.7]{a}
\node[left,scale=#5*1.5] at (-0.7,-0.7){$#1$};
\Vertex[x=0.7,y=-0.7]{b}
\node[right,scale=#5*1.5] at (0.7,-0.7){$#2$};
\Vertex[x=0.7,y=0.7]{c}
\node[right,scale=#5*1.5] at (0.7,0.7){$#3$};
\Vertex[x=-0.7,y=0.7]{d}
\node[left,scale=#5*1.5] at (-0.7,0.7){$#4$};
\Edges(a,b,c,d,a)
\end{scope}
\end{tikzpicture}
}

\begin{document}

\title{\bf  Open Quantum Symmetric Simple Exclusion Process }

\author{Denis Bernard}
\author{Tony Jin}
\affiliation{Laboratoire de Physique de l'Ecole Normale Sup\'erieure de Paris, CNRS, ENS $\&$ Universit\'e PSL, Sorbonne Universit\'e, Universit\'e Paris Diderot, France}

\date{\today}

\begin{abstract} 
We present the solution to a model of fermions hopping between neighbouring sites on a line with random Brownian amplitudes and open boundary conditions driving the system out of equilibrium. The average dynamics reduces to that of the symmetric simple exclusion process. However, the full distribution encodes for a richer behaviour entailing fluctuating quantum coherences which survive in the steady limit. We determine exactly the steady statistical distribution of the system state. We show that the out of equilibrium quantum coherence fluctuations satisfy a large deviation principle and we present a method to recursively compute exactly the large deviation function. As a by-product, our approach gives a solution of the classical symmetric simple exclusion process based on fermion technology. Our results open the route towards the extension of the macroscopic fluctuation theory to many body quantum systems.
\end{abstract}

\maketitle

{\bf {Introduction}.--} Non-equilibrium phenomena are ubiquitous in Nature. Understanding the fluctuations of the flux of heat or particles through systems is a central question in non equilibrium statistical mechanics. Last decade has witnessed tremendous conceptual and technical progresses in this direction for classical systems, starting from the exact analysis of simple models~\cite{Kipnis99,Liggett99,Spohn91}, such as the Symmetric Simple Exclusion Process (SSEP)~\cite{SSEP,Eyink91,Derrida_Review,Mallick_Review}, via the understanding of fluctuation relations~\cite{Fluctu95,Jar97,Crooks99} and their interplay with time reversal~\cite{Maes99,Maes_bis}, and culminating in the formulation of the macroscopic fluctuation theory (MFT) which is an effective theory adapted to describe transport and its fluctuations in diffusive classical systems~\cite{MFT}.  Whether MFT may be extended to the quantum realm is yet unexplored. 

In parallel, the study of quantum systems out of equilibrium has received a large amount of attention in recent years~\cite{Review-Q1,Review-Q2,Review-Q3,Review-Q4}. Experimentally, unprecedented control of cold atom gases gave access to the observation of many body quantum systems in inhomogeneous and isolated setups~\cite{cold_Hild14,cold_Boll16,cold_Bouchoule17,cold_Tang18,cold_Rauer18}. Theoretically, results about closed, quantum systems  have recently flourished, with a better perception of the roles of integrability, chaos or disorder~\cite{Qclosed_R1,Qclosed_R2,Qclosed_R3,Qclosed_R4,Qclosed_R5,Qclosed_ter,Qclosed_R6,Qclosed_R7,Qclosed_R8,Chaos_Bound16,Qclosed_R9,Prosen18}. In critical or integrable models, a good understanding has been obtained with a precise description of entanglement dynamics, quenched dynamics, as well as transport~\cite{Entangl_1,Entangl_2,Entangl_3,Entangl_4,Entangl_5,Quench_1,BD1,BD2,Moore12,Prosen_bis,GHDiff18,Prosen_ter}. These efforts culminated in the development of a hydrodynamic picture adapted to integrable systems~\cite{GHD1,GHD2}. However, these understandings are restricted to closed, predominantly ballistic, systems.

Many quantum transport processes are diffusive rather than ballistic~\cite{Hartnoll15} and, to some extends, physical systems are generically in contact with external environments. It is thus crucial to extend the previous studies by developing simple models for fluctuations in open, quantum many body, locally diffusive, out of equilibrium systems. Putting aside the quantum nature of the environments leads to consider model systems interacting with classical reservoirs or noisy external fields. In the context of quantum many body systems, and especially quantum spin chains, the study of such models has recently been revitalised \cite{Znid10,BBen,Garrah17,Eisert17,Knap18,Lama18,Xu18,Gullans18}, partly in connection with random quantum circuit \cite{randomU1,randomU2,randomU3,Qrandom-circuit1,Qrandom-circuit2,Qrandom-circuit3,Gullans19}, as a way to get a better understanding of entanglement production or information spreading.

In this work, we introduce and solve an iconic example of such models. It is a stochastic variant of the Heisenberg XX spin chain. It codes for typical features of quantum many body at scales smaller than the coherence length (to keep interference effects) but larger than the mean free path (to include diffusion). It describes fermions hopping from site to site on a discretised line, but with Brownian hopping amplitudes, and interacting with reservoirs at the chain boundaries.  For reasons explained below, we may call this model the {\it quantum SSEP}. Its average dynamics reduces to the classical SSEP, but the model codes for the fluctuations around this mean behaviour. Although decoherence is at play in the mean behaviour, fluctuating quantum coherences survive to the noisy interaction. Their magnitudes typically scale proportionally with the inverse of the square root of the system size. We characterise completely the steady measure on the system state which encodes for the fluctuations of the quantum coherences and occupation numbers at large time. We also present a recursive method to compute exactly, order by order, the large deviation function of these fluctuations. These findings open the route towards the extension of the MFT~\cite{MFT} to many body quantum systems.   

{\bf {The open quantum SSEP}.--}
For an open chain in contact with external reservoirs at their boundaries, the quantum SSEP dynamics results from the interplay between unitary, but stochastic, bulk flows with dissipative, but deterministic, boundary couplings. The bulk flows induce unitary evolutions of the system density matrix $\rho_t$ onto $e^{-idH_t}\,\rho_t\,e^{idH_t}$ with Hamiltonian increments
\beq \label{eq:defHXXsto}
dH_{t}=\sqrt{D}\,\sum_{j=0}^{L-1}\big(c_{j+1}^{\dagger}c_{j}\,dW_{t}^{j}+c_{j}^{\dagger}c_{j+1}\,d\overline W_{t}^{j}\big),
\eeq
for a chain of length $L$, where $c_{j}$ and $c_{j}^{\dagger}$ are canonical fermionic operators, one pair for each site of the chain, with $\{c_{j},c_{k}^{\dagger}\}=\delta_{j;k}$, and $W_{t}^{j}$ and  $\overline W_{t}^{j}$ are pairs of complex conjugated Brownian motions, one pair for each edge along the chain, with quadratic variations $dW_{t}^{j}d\overline{W}_{t}^{k}=\delta^{j;k}\,dt$. This model was shown to describe the effective dynamics of the stochastic Heisenberg XX spin chain with dephasing noise in the strong noise limit~\cite{BBJsto}. It codes for a diffusive evolution of the number operators $\hat n_{j}=c_{j}^{\dagger}c_{j}$ with the parameter $D$ being the diffusion constant. This model is one of the simplest model of quantum, stochastic, diffusion. It shares similarities with that of \cite{Gullans18}. Exact results concerning the statistical mean behaviour of this model, and more generally of the dephasing Heisenberg spin chain, were described in \cite{Znidaric10a,Znidaric11,Eisler11,Essler-Pro16}.
Properties of the closed periodic version of this model were deciphered in~\cite{BBJbis} via a mapping to random matrix theory. We set $D=1$ in the following.

Assuming the interaction between the chain and the reservoirs to be Markovian, the contacts with the external leads can be modelled by Lindblad terms~\cite{Lindblad}. The resulting equations of motion read
\beq \label{eq:Q-flow}
 d\rho_{t}=-i[dH_{t},\rho_{t}] -\frac{1}{2}[dH_{t},[dH_{t},\rho_{t}]]+\mathcal{L}_\mathrm{bdry}(\rho_{t})dt,
 \eeq
with $dH_t$  as above and $\mathcal{L}_\mathrm{bdry}$ the boundary Lindbladian. The two first terms result from expanding the unitary increment $\rho_t \to e^{-idH_t}\,\rho_t\,e^{idH_t}$ to second order (because the Brownian increments scale as $\sqrt{dt}$). The third term codes for the dissipative boundary dynamics
with $\mathcal{L}_\mathrm{bdry}=\alpha_{0}{\cal L}_{0}^{+}+\beta_{0}{\cal L}_{0}^{-}+\alpha_{L}{\cal L}_{L}^{+}+\beta_{L}{\cal L}_{L}^{-}$ and 
\begin{align} \label{eq:L-bdry}
{\cal L}_{j}^{+}(\bullet) & =c_{j}^{\dagger}\bullet c_{j}-\frac{1}{2}(c_{j}c_{j}^{+}\bullet+\bullet c_{j}c_{j}^{\dagger}),\\
{\cal L}_{j}^{-}(\bullet) & =c_{j}\bullet c_{j}^{\dagger}-\frac{1}{2}(c_{j}^{\dagger}c_{j}\bullet+\bullet c_{j}^{\dagger}c_{j}),
\end{align}
where the parameters $\alpha_j$ (resp. $\beta_j$) are the injection (resp. extraction) rates.

The dynamics being noisy, so is the density matrix and hence the quantum expectations such as the mean quantum  occupation numbers $n_j=\Tr(\hat n_j\, \rho_t)$. Their stochastic averages $\mathbb{E}[n_j]$ evolve according to 
\[ \partial_t\mathbb{E}[n_j]= \Delta^\mathrm{dis}_j\mathbb{E}[n_j] + \hskip -0.3truecm \sum_{k\in\{0,L\}}\hskip -0.2truecm \delta_{j;k}\big(\alpha_{k}(1-\mathbb{E}[n_k])-\beta_{k}\mathbb{E}[n_k]\big),\]
with $\Delta^\mathrm{dis}_j$ the discrete Laplacian, $\Delta^\mathrm{dis}_j n_j=n_{j+1}-2n_j+n_{j-1}$, illustrating the diffusive bulk dynamics and the boundary injection/extraction processes. At large time, they reach a linear profile,
\beq
n_j^*:= \lim_{t\to\infty} \mathbb{E}[n_j] = \frac{n_a(L+b-j) + n_b(j+a)}{L+a+b},
\eeq
with $n_{a}:=\frac{\alpha_{0}}{\alpha_{0}+\beta_{0}}$, $n_{b}:=\frac{\alpha_{L}}{\alpha_{L}+\beta_{L}}$,
$a:=\frac{1}{\alpha_{0}+\beta_{0}}$, $b:=\frac{1}{\alpha_{L}+\beta_{L}}$, associated to a steady mean flow from one reservoir to the other. In the large size limit, $L\to\infty$ at $x=i/L$ fixed, this mean profile, $n^*(x)= n_a + x(n_b-n_a)$, interpolates linearly the two boundary mean occupations $n_a$ and $n_b$,  in agreement with \cite{Znidaric10a,Znidaric11,Eisler11,Essler-Pro16,Gullans18}.

In mean, the off-diagonal quantum expectations $G_{ji}:=\Tr(c_i^\dag c_j\, \rho_t)$ vanish exponentially fast, $\lim_{t\to\infty}\mathbb{E}[G_{ji}]=0$ for $j\not=i$, hence reflecting decoherence due to destructive interferences induced by the noise. However, this statement is only valid in mean as fluctuating coherences survive at sub-leading orders with a rich statistical structure, with long range correlations.

{\bf {The steady state~: fluctuations and coherences}.--}
As exemplified by the above one-point functions, a steady state is attained at large time in the sense that the distribution of quantum expectations reaches a stationary value. Equivalently, the limit $ \lim_{t\to\infty} \mathbb{E}[F(G_t)]$ exists for any smooth function $F$ of the matrix of two-point quantum expectations $G$ and this defines an invariant measure $\mathbb{E}_\infty[\bullet]$ of the flow (\ref{eq:Q-flow}), that we shall denote by $[\bullet]$ to simplify the notation. Diagonal elements $G_{jj}$ code for occupation numbers while the off-diagonal elements $G_{ji}$ for coherences, and hence $[\bullet]$ for their steady statistics.

Amongst the two point functions $\mathbb{E}[G_{ij}G_{kl}]$, only those with $\{i=j,k=l\}$ and $\{i=l,j=k\}$ survive at large time, the others decrease exponentially fast. This leaves us with three possible configurations: $[G_{ii}^2]$, $[G_{ii}G_{jj}]$, and $[G_{ij}G_{ji}]$, $j\not=i$, coding respectively for quantum occupation and coherence fluctuations. They are determined by solving the stationarity equations for the invariant measure (see Supplementary Material):
\begin{align*} 
 [G_{ij} G_{ji}]^c&= \frac{(\Delta n)^2\, (i+a)(L-j+b)}{(L+a+b-1)(a+b+L)(a+b+L+1)},\\
[G_{ii}G_{jj}]^c &=- \frac{(\Delta n)^2\, (i+a)(L-j+b)}{(L+a+b-1)(a+b+L)^2(a+b+L+1)},\\
[G_{ii}^2]^c &= \frac{(\Delta n)^2\, \big(2(i+a)(L-i+b)-(L+a+b)\big)}{2(a+b+L)^2(a+b+L+1)},
\end{align*}
for $i<j$ with $\Delta n=n_b-n_a$ and $ [G_{ii} G_{jj}]^c=[G_{ii}G_{jj}]-[G_{ii}][G_{jj}]$. The first lesson is that coherences are present in the large time steady state as their covariances do not vanish exponentially but remains finite. At large size, $L\to\infty$ with $x=i/L$, $y=j/L$ fixed, their second moments behave as
\begin{align} \label{eq:N=2-QC}
 [G_{ij} G_{ji}]^c&= \frac{1}{L}(\Delta n)^2\, x(1-y) +O(L^{-2}),\\
[G_{ii}G_{jj}]^c &=- \frac{1}{L^2}(\Delta n)^2\, x(1-y) +O(L^{-3}), 
\end{align}
for $x<y$, while $[G_{ii}^2]^c =  \frac{1}{L}(\Delta n)^2\, x(1-x) +O(L^{-2})$. 
The second lesson is, on one hand, that these fluctuating coherences scale as $1/\sqrt{L}$ in the thermodynamic limit, and in the other hand, that the correlations between the quantum occupation numbers $n_i$ and $n_j$ at distinct sites $i\not=j$ scale as $1/L^2$ and hence are sub-leading. These correlations coincide with those of the statistical mean of the number operator two-point expectations, for reasons explained below, but this coincidence does not hold for higher ($N>3)$ point correlations.

These facts hold for higher order cumulants $[G_{i_1j_1}\cdots G_{i_Nj_N}]^c$ of the matrix of two-point quantum expectations. These cumulants are non vanishing only if the sets $\{i_1,\cdots,i_N\}$ and $\{j_1,\cdots,j_N\}$ coincide so that the $N$-uplet $(j_1,\cdots,j_N)$ is a permutation of $(i_1,\cdots,i_N)$. To such product $G_{i_1j_1}\cdots G_{i_Nj_N}$ we can associate an oriented graph with a vertex for each point $i_1,\cdots,i_N$ and an oriented edge from $i$ to $j$ for each insertion of $G_{ji}$. These graphs may be disconnected. The condition that the sets $\{i_1,\cdots,i_N\}$ and $\{j_1,\cdots,j_N\}$ coincide translates into the fact that the number of ongoing edges equals that of outgoing edges, at each vertex. 
For instance, $[G_{ii}]$ is represented by $[\hspace{3pt} \graphN{i}{0.5} \hspace{2pt} ]$, $[G_{ii}G_{jj}]$ for $i \not= j$ by $[\hspace{3pt}\graphNcroixdeux{i}{j}{0.5}\hspace{2pt}]$, $[G_{ij}G_{ji}]$ for $i\not= j$ by $[\graphFdeux{i}{j}{0.5}]$ and $[G_{ii}^{2}]$ by [$\graphNdeux{i}{0.5}]$.
 
The claim is that expectations of single loop diagrammes, corresponding to the expectations of cyclic products $[G_{i_1i_N}\cdots G_{i_3i_2}G_{i_2i_1}]^c$, are the elementary building blocks in the large size limit. They scale proportionally to $1/L^{N-1}$ in the thermodynamic limit
\beq \label{eq:Q-loop}
[G_{i_1i_N}\cdots G_{i_3i_2}G_{i_2i_1}]^c =\frac{1}{L^{N-1}}\, g_N(x_1,\cdots,x_N) + O(L^{-N}),
\eeq
with $x_p=i_p/L$. The expectations $g_N$ depend on which sector the points ${\bm x}:=(x_1,\cdots,x_N)$ belong to, with the sectors indexing how the ordering of the points along the chain match / un-match that along the loop graph.
Let us choose to fix an ordering of the points along the chain so that $0\leq x_1<\cdots<x_N \leq1$, and let $\sigma$ be the permutation coding for the ordering of the point vertices around the loop so that by turning around the oriented loop one successively encounters the vertices labeled by $x_{\sigma(1)}$, $x_{\sigma(2)}$, $\cdots$, up to $x_{\sigma(N)}$. There are $(N-1)!/2$ sectors because the ordering around the loop is defined up to cyclic permutations and because reversing the orientation of the loop preserves the expectations. Let us then set $f_N^\sigma({\bm x}):=g_N(x_{\sigma(1)}, \cdots, x_{\sigma(N)})$. 

\begin{figure}[t] 
\includegraphics[width=0.45\textwidth]{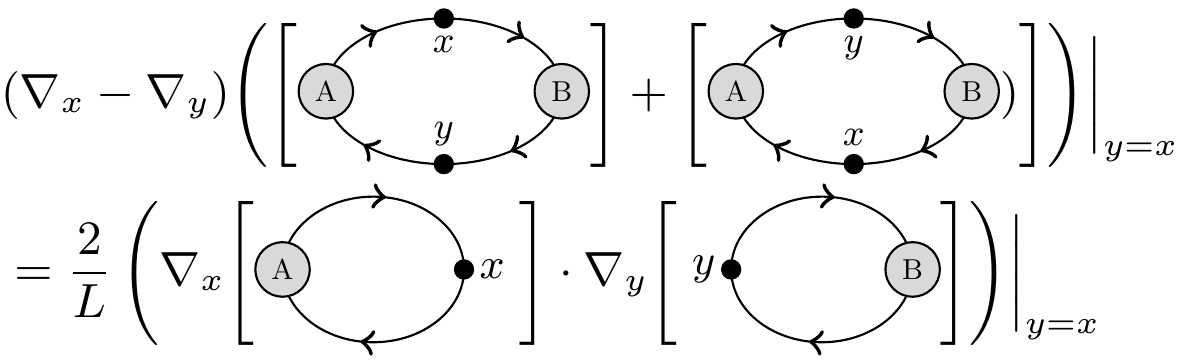}
\caption{Graphical representation of the contact relation (\ref{eq:Q-Neuman}).}
  \label{fig:contact-relation}
\end{figure}

The $f_N^\sigma$'s are recursively determined by a set of equations which arise from the stationarity conditions of the invariant measure. (See the Supplementary Material). First, stationarity in the bulk imposes that $\Delta_{x_j}\,f_N^\sigma({\bm x})=0$ for all $j$ with $\Delta_{x}$ the Laplacian with respect to $x$, as a consequence of the bulk diffusivity. Second, the couplings at the boundaries freeze the fluctuations so that 
\beq \label{eq:Q-bdry}
f_N^\sigma({\bm x})\vert_{x_1=0}=f_N^\sigma({\bm x})\vert_{x_N=1}=0.
\eeq
Third, contact interactions due to noisy hoppings imposes two conditions on expectations at the boundary between the sectors $\sigma$ and $\pi_{j;j+1}\sigma$ with $\pi_{j;j+1}$ the permutation transposing $j$ and $j+1$. The ordering of the point vertices in the sector $\sigma$ and $\pi_{j;j+1}\sigma$ differ by the exchange of $x_j$ and $x_{j+1}$, so that $x_{j+1}=x_j$ at these boundaries. The first contact condition is the continuity condition
\beq \label{eq:Q-cont}
f_N^\sigma({\bm x})\vert_{x_{j+1} = x_j}=f_N^{\pi_{j;j+1}\sigma}({\bm x})\vert_{x_{j+1}=x_j} .
\eeq 
To write the second contact condition, let us define $j_*^-$ (resp. $j_*^+$) to be the $\sigma$ pre-image of $j$ (resp. $j+1$), i.e. $j=\sigma(j_*^-)$ and $j+1=\sigma(j_*^+)$. Since, the vertices $x_{j+1}$ and $x_{j}$ are identified at these sector boundaries, the loop graph splits into two sub-loops, touching at the vertex $x_j$, one including the circle arc $x_{\sigma(j_*^--1)},x_j,x_{\sigma(j_*^++1)}$, and the other containing the circle arc $x_{\sigma(j_*^+-1)},x_j,x_{\sigma(j_*^-+1)}$, respectively denoted $\ell^{\sigma,-}_j$ and $\ell^{\sigma,+}_{j}$. The second contact condition is the Neumann like matching condition
\begin{align} \label{eq:Q-Neuman}
&(\nabla_{x_{j}}-\nabla_{x_{j+1}})\big(f_N^\sigma({\bm x})+f_N^{\pi_{j;j+1}\sigma}({\bm x})\big)\vert_{x_{j+1}=x_j} \\
=&\, 2 \, \big(\nabla_{x_{j}} [\mathfrak{R}^+_j\cdot f^\sigma]({\bm x})\big) \cdot \big(\nabla_{x_{j}}  [\mathfrak{R}^-_{j}\cdot f^\sigma]({\bm x})\big), \nonumber
\end{align}
with $[\mathfrak{R}^\pm_j\cdot f^\sigma]$ the expectations of the reduced sub-loops $\ell^{\sigma,\pm}_j$. Eqs. (\ref{eq:Q-bdry},\ref{eq:Q-cont},\ref{eq:Q-Neuman}) allow to recursively compute the building block loop expectations (\ref{eq:Q-loop}).
See FIG.~\ref{fig:contact-relation} for a graphical representation of (\ref{eq:Q-Neuman}).

Furthermore, connected expectations of pinched graphs obtained by identifying points in single loop graphs are obtained by continuity from the expectations of the corresponding parent loop graphs, thanks to (\ref{eq:Q-cont}). They are of order $1/L^{N-1}$ with $N$ the number of edges in the pinched graph (and hence the number of insertions of matrix elements of $G$). All other connected expectations of disconnected graphs are sub-leading in the large size limit.

The conditions (\ref{eq:Q-bdry},\ref{eq:Q-cont},\ref{eq:Q-Neuman}) allow to determine all leading expectations recursively. For $N=3$, there is only one sector and $g_3(x,y,z)= (\Delta n)^3\, x(1-2y)(1-z)$ for $x<y<z$, so that
\beq
[G_{ik}G_{kj}G_{ij}]^c= \frac{1}{L^2} (\Delta n)^3\, x(1-2y)(1-z) +O(L^{-3}),
\eeq
with $x=i/L$, $y=j/L$ and $z=k/L$ ($i<j<k$). For $N=4$, there are $3$ sectors respectively associated to the identity and the transpositions $\pi_{1;2}$ and $\pi_{2;3}$ :
\begin{equation*}
\hspace{10pt} \graphFquatre{x_1}{x_2}{x_3}{x_4}{0.5}\hspace{10pt}, \hspace{10pt} \graphFquatre{x_2}{x_1}{x_3}{x_4}{0.5}\hspace{10pt}, \hspace{10pt} \graphFquatre{x_1}{x_3}{x_2}{x_4}{0.5} \hspace{15pt}.
\end{equation*}
 For $x_1<x_2<x_3<x_4$, their expectations are respectively :
\begin{align*}
\frac{1}{L^3}(\Delta n)^4\, x_1(1-3x_2-2x_3+5x_2x_3)(1-x_4),\\
\frac{1}{L^3}(\Delta n)^4\, x_1(1-3x_2-2x_3+5x_2x_3)(1-x_4),\\
\frac{1}{L^3}(\Delta n)^4\, x_1(1-4x_2-x_3+5x_2x_3)(1-x_4),
\end{align*}
up to $O(L^{-4})$ contributions.

The scaling behaviour of the single loop expectations (\ref{eq:Q-loop}) ensures that the fluctuations of the matrix of quantum two-point expectations $G$ satisfy a large deviation principle, in the sense that their generating function is such that $\big[e^{\Tr(AG)}\big]\asymp_{L\to\infty} e^{L\,\mathfrak{F}(A)}$ for some function $\mathfrak{F}(A)$, called the large deviation function, 
\beq
\mathfrak{F}(A)=\lim_{L\to\infty} \frac{1}{L}\log \left[ e^{\Tr(AG)}\right].
\eeq
This function admits a series expansion, $\mathfrak{F}(A)=\sum_N \frac{1}{N!}\, \mathfrak{F}^{(N)}$, with the $\mathfrak{F}^{(N)}$'s given by the multiple sums $L^{-N}\sum_{i_1,\cdots, i_N} [G_{i_1i_N}\cdots G_{i_3i_2}G_{i_2i_1}]^c\, (A_{i_1i_2}\cdots A_{i_Ni_1})$ which converge to multiple integrals.
To lowest order
\begin{align}
\mathfrak{F}(A) &= \int_0^1 \hskip -0.15 truecm dx\,  n^*(x)A(x,x)  \\
+\, {(\Delta n)^2} & \int_0^1 \hskip -0.15 truecm dx \hskip -0.05 truecm  \int_x^1 \hskip -0.15 truecm dy\ x(1-y)\, A(x,y)A(y,x) +\cdots .
\nonumber
\end{align}
Higher orders can be recursively computed by using equations (\ref{eq:Q-bdry},\ref{eq:Q-cont},\ref{eq:Q-Neuman}).

\begin{figure}[t]  
\includegraphics[width=0.4\textwidth]{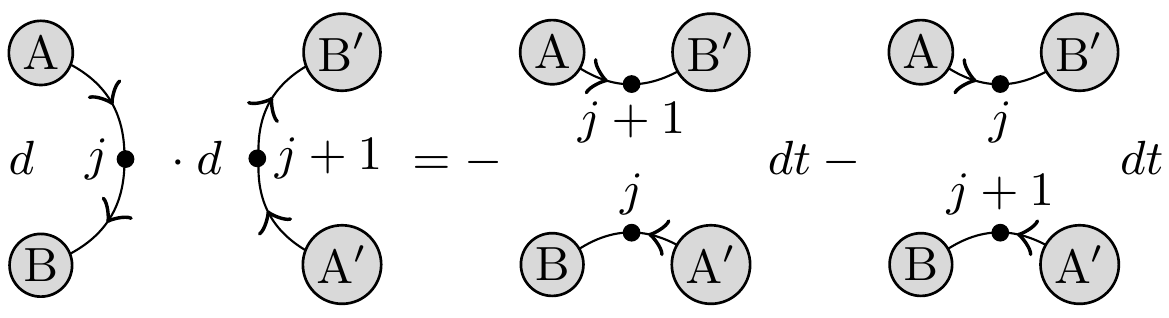}
\caption{Graphical representation of the reshuffling relation.}
 \label{fig:reshuffling-relation}
\end{figure}

{\bf {Sketch of proof}.--} 
Since both the Hamiltonian increments (\ref{eq:defHXXsto}) and the Lindbladians (\ref {eq:L-bdry}) are quadratic in the fermionic creation and annihilation operators, the  stochastic evolution (\ref{eq:Q-flow}) preserves Gaussian states of the form $\rho_{t}=Z_t^{-1}\, e^{c^{\dagger}M_{t}c}$ with $M_t$ a $L\times L $ matrix and $Z_t=\mathrm{Tr}(e^{c^{\dagger}M_{t}c})$. These density matrices are parametrised by $M_t$ or, equivalently, by the matrix of quantum two-point expectations $G_{ij}=\mathrm{Tr}(\rho_t c_{j}^{\dagger}c_{i})$. One can show that $G_{t}=\frac{e^{M_{t}}}{1+e^{M_{t}}}$. Eq. (\ref{eq:Q-flow}) then becomes a stochastic equation for $M_t$ or $G_t$. For instance, for $0\not=i<j\not=L$,
\begin{align} \label{eq:stoQ}
dG_{ij} & =-2\text{\,}G_{ij}dt +i\big(G_{i;j-1}d\overline{W}_{t}^{j-1}+G_{i;j+1}dW_{t}^{j}) \nonumber\\
& \hspace{-1em}-i\big(G_{i-1;j}dW_{t}^{i-1}+G_{i+1;j}d\overline{W}_{t}^{i}\big),
\end{align}
with similar equations for $G_{ii}$ and at the two chain boundaries. Imposing the stationarity of the measure amounts to demand that the statistical expectations $[F(G_t)]$ are time independent for any function $F$. Since the It\^o derivatives of polynomials in $G_t$ are polynomials in $G_t$ of the same degrees, the stationarity conditions are sets of linear equations on moments of given order. There are two types of contributions arising from the It\^o derivatives of polynomials: one completing the drift term in (\ref{eq:stoQ}) to produce discrete Laplacians acting on products of $G_t$'s, the other producing contact interactions. For instance, $dG_{kj}dG_{l;j+1}\vert_\mathrm{contact}=-(G_{k;j+1}dW^j)(G_{lj}d\overline W^j)= - G_{k,j+1}G_{lj}\,dt$ which implements the transposition of the adjacent points $j$ and $j+1$. As a consequence, the It\^o derivatives of graphs coding for products of $G_t$'s with adjacent indices induce a reshuffling of the connections of these graphs. See FIG.~\ref{fig:reshuffling-relation} for an illustration. Thus, the stationarity conditions yield relations between expectations of reshuffled graphs from which the relations (\ref{eq:Q-bdry},\ref{eq:Q-cont},\ref{eq:Q-Neuman}) can be deduced. (See Supplementary Material). More details will be described elsewhere~\cite{Qssep_long}.

{\bf {Connecting to the classical SSEP}.--}
The mean density matrix $\bar \rho_t:=\mathbb{E}[\rho_t]$ evolves according to the Lindblad equation $\partial_t\bar \rho_t=\mathcal{L}_\mathrm{bulk}(\bar \rho_t) + \mathcal{L}_\mathrm{bdry}(\bar \rho_t)$ with $\mathcal{L}_\mathrm{bdry}$ defined in (\ref{eq:L-bdry}) and bulk Lindbladian
\beq \label{eq:Lbulk}
\mathcal{L}_\mathrm{bulk}(\bar \rho_t)= -\frac{1}{2}\sum_{j=0}^{L-1}\big( [c_{j+1}^\dag c_j,[c_j^\dag c_{j+1},\bar \rho_t]]+\mathrm{h.c}\big).
\eeq
This Lindblad dynamics has been studied in \cite{Znidaric10a,Znidaric11,Eisler11,Essler-Pro16}.
For density matrices diagonal in the occupation number basis, it codes for the time evolution of SSEP.
Asymptotically in time, decoherence is effective and the mean density matrix is diagonal, $\bar \rho_t = \sum_{[n]} \bar Q_t[\mathfrak{n}]\, \mathbb{P}_{[\mathfrak{n}]}$ where the $\mathbb{P}_{[\mathfrak{n}]}$'s are the projectors on the occupation number eigen-states $|{[\mathfrak{n}]}\rangle$ and $\bar Q_t[\mathfrak{n}]$ the mean populations. The $\mathbb{P}_{[\mathfrak{n}]}$'s are products of projectors $\mathbb{P}_j^{\mathfrak{n}_j}$ on each site of the chain, with $\mathfrak{n}_j=0$ (resp. $\mathfrak{n}_j=1$) for empty (resp. full). On adjacent pairs of projectors, the bulk Lindbladian acts as
\begin{align*}
\mathcal{L}_\mathrm{bulk}(\mathbb{P}_j^{1}\mathbb{P}_{j+1}^{0}) &= -\mathbb{P}_j^{1}\mathbb{P}_{j+1}^{0} + \mathbb{P}_j^{0}\mathbb{P}_{j+1}^{1},\\
\mathcal{L}_\mathrm{bulk}(\mathbb{P}_j^{0}\mathbb{P}_{j+1}^{1}) &= - \mathbb{P}_j^{0}\mathbb{P}_{j+1}^{1} + \mathbb{P}_j^{1}\mathbb{P}_{j+1}^{0},
\end{align*}
whereas $\mathcal{L}_\mathrm{bulk}(\mathbb{P}_j^{0}\mathbb{P}_{j+1}^{0})=0$ and $\mathcal{L}_\mathrm{bulk}(\mathbb{P}_j^{1}\mathbb{P}_{j+1}^{1})= 0$. This is equivalent to the SSEP transition matrix. 

As a consequence, the SSEP generating function for the occupation number fluctuations can be identified with the statistical average of the generating function of quantum expectations of the number operators,
\beq \nonumber
 \langle e^{\sum_j a_j \mathfrak{n}_j}\rangle_\mathrm{ssep} = \mathrm{Tr}\big( \bar \rho\, e^{\sum_i a_i \hat n_i}\big)
= \Big[ \mathrm{Tr}\big( \rho\, e^{\sum_i a_i \hat n_i}\big)\Big] ,
\eeq
with $\hat n_i=c^\dag_ic_i$. It can be computed using Wick's theorem, so that the SSEP cumulants read
\beq
\vev{\mathfrak{n}_{j_1}\cdots \mathfrak{n}_{j_N}}^c_\mathrm{ssep}= \frac{(-)^{N-1}}{L^{N-1}} \sum_\sigma f_N^\sigma({\bm x}) +O(L^{-N}) ,
\eeq
with  $x_k=j_k/L$ all distinct. The sum is over permutations $\sigma$ modulo cyclic permutations. (See Supplementary Material). The expectations (\ref{eq:Q-loop}) of the matrix of quantum two-point expectations cannot be reconstructed from the SSEP expectations (for $N\geq4$)  because the latter are symmetric under permutations and hence only involve the sum of the sectors.

{\bf {Discussion}.--} 
We have introduced a quantum extension of the SSEP and outlined how to solve it exactly by characterising its invariant measure and computing the large deviation function of the matrix of quantum two-point expectations. The quantum SSEP is a simple, if not the simplest, model coding for diffusive behaviour of quantum operators in a many body fermionic systems. In mean, it reduces to the classical SSEP so that the statistical averages of the quantum expectations of the number operators coincide with those of the classical SSEP. 

The quantum SSEP is strictly finer than its classical counterpart, and contains much more information, including fluctuations of quantum coherences. Although decoherence is at work on the mean steady state, we have observed and quantified sub-leading fluctuating coherences which are not visible in the mean behaviour~\cite{Znidaric10a,Znidaric11,Eisler11}. In the thermodynamic large size limit, the system state approaches a self averaging non equilibrium state dressed by occupancy and coherence fluctuations whose amplitudes scale proportionally to $1/\sqrt{L}$. We have described how to compute the large deviation function for these fluctuations, order by order. One simple experimental route to probe these coherences consists in transposing to our system the recently proposed setup \cite{Gullans18} to conduct interferometry experiments between two parts of the system.
Another possibility to generate echoes of coherence effects in the occupancy number correlations consists in injecting fermions in quantum states which are not eigenstates of the occupancy operators.

As an example of quantum out of equilibrium exclusion processes and of fluctuating quantum discrete hydrodynamics, our findings open several new research directions. The first concerns the integrable structure underlying the exact solution we have presented and its connection with the existing solution methods for classical exclusion processes~\cite{Derrida93,deGier05,Evans07,Derrida07}. The second concerns the extension of our work to deal with the quantum analogue of the asymmetric simple exclusion process (ASEP). We have already noticed that the appropriate generalisation amounts to couple the fermionic system to quantum noise~\cite{Qnoise1,Qnoise2}. But the most important ones deal with using the present model, and its generalisations to interacting systems, to formalise the extension of the MFT~\cite{MFT} to many body quantum systems. Proposing such quantisation of the MFT incorporating the fluctuating quantum coherences requires going beyond the statistical mean behaviour.
It has been observed in \cite{Znidaric14} that the additivity principle~\cite{Additivity_MFT}, which applies classically with some degree of generality, also  holds for some statistics encoded into the mean system state of diffusive spin chains. How to extend this principle to keep track of the quantum coherences and their statistical fluctuations? How to take the continuum limit of those models to provide a quantisation of the MFT?
We plan to report on these questions in the near future~\cite{Prepa}.

{\bf {Acknowledgments}.--} 
Both authors wish to acknowledge Michel Bauer for discussions and past and future collaborations. 
D.B. acknowledges support from the CNRS and the ANR via the contract ANR-14-CE25-0003.


\newpage
\appendix
\onecolumngrid

\vspace{1cm}
\begin{center}
\textbf{{\Large Supplemental Material}}
\end{center}
\begin{center}
\textbf{{\large  Open Quantum Symmetric Simple Exclusion Process}}
\end{center}
\begin{center}
{Denis Bernard and Tony Jin}
\end{center}

\section{Low order cumulants}

From eq.(2) in the main text, we derive the stochastic equations satisfied by the matrix $G_t$:
\begin{align}
dG_{i;i} & =(G_{i+1;i+1}-2G_{i;i}+G_{i-1;i-1})dt+\sum_{p\in\{0,L\}}\delta_{i,p}(\alpha_{p}(1-G_{i;i})-\beta_{p}G_{i;i})dt \label{eq:Main_sto1}\\
 & \hspace{1em}+i\big(G_{i;i-1}d\bar{W}_{t}^{i-1}+G_{i;i+1}dW_{t}^{i}-G_{i-1;i}dW_{i-1}^{t}-G_{i+1;i}d\bar{W}_{t}^{i}\big), \nonumber\quad(j=i)\\
dG_{i;j} & =-2\text{\,}G_{i;j}dt-\frac{1}{2}\sum_{p\in\{0,L\}}(\alpha_{p}+\beta_{p})(\delta_{i,p}+\delta_{j,p})G_{i;j}dt  \label{eq:Main_sto2} \\
 & \hspace{1em}+i\big(G_{i;j-1}d\bar{W}_{t}^{j-1}+G_{i;j+1}dW_{t}^{j}-G_{i-1;j}dW_{t}^{i-1}-G_{i+1;j}d\bar{W}_{t}^{i}\big),\text{\quad}(j\not=i)\nonumber 
\end{align}
Here and in the following, it will always be implicitly assumed that we have to truncate the equations keeping only the appropriate terms when evaluating them at boundaries, i.e. either taking $i$ or $j$ equal to $0$ or $L$. For instance 
\begin{align} \label{eq:Main_sto3}
dG_{0;j} & =-G_{0;j}dt-\frac{1}{2}\sum_{p\in\{0,L\}}(\alpha_{p}+\beta_{p})(\delta_{0,p}+\delta_{j,p})G_{0;j}dt\\
 & \hspace{1em}+i\big(G_{0;j-1}d\bar{W}_{t}^{j-1}+G_{0;j+1}dW_{t}^{j}-G_{1;j}d\bar{W}_{t}^{0}\big),\text{\quad}(j\not=0), \nonumber
\end{align}
and similarly at the other end of the chain.
\bigskip 

{\bf Cumulants of order 2}.--
Let us see how to compute the  cumulants of order $2$.
To each product of $G$'s not vanishing in the long-time limit, there is an associated stationary equation. Let's illustrate how this works for instance for $\mathbb{E}[G_{i;j}G_{j;i}]$ with $i$ and $j$ $\neq$ $0$ or $L$. Working in the It\^o convention, the differential of $\mathbb{E}[G_{ij}G_{ji}]$ is given by~: 
\begin{align}
d\mathbb{E}[G_{i;j}G_{j;i}] & =\mathbb{E}[dG_{i;j}G_{j;i}+dG_{i;j}G_{j;i}+dG_{i;j}dG_{j;i}] \nonumber \\
 & =\mathbb{E}[((\Delta_{i}^{\mathrm{dis}}+\Delta_{j}^{\mathrm{dis}})G_{i;j}G_{j;i}+2(\delta_{i,j}-\delta_{i-1,j})G_{i-1;i-1}G_{i;i}+2(\delta_{i,j}-\delta_{i+1,j})G_{i;i}G_{i+1;i+1}) \nonumber \\
 & \hspace{1em}-\sum_{p\in\{0,L\}}\gamma_{p}(\delta_{i,p}+\delta_{j,p})(G_{i;j}G_{j;i})]dt,
\end{align}
where $\Delta_{i}^\mathrm{dis}$ is the discrete Laplacian as in the main text. To go from the first line to the second line, we made use of
the statistical properties of the complex noises, i.e $\mathbb{E}[dW_{t}^{j}]=\mathbb{E}[d\overline{W}_{t}^{j}]=\mathbb{E}[(dW_{t}^{j})^{2}]=\mathbb{E}[(d\overline{W}_{t}^{j})^{2}]=0$
and $\mathbb{E}[dW_{t}^{i}d\overline{W}_{t}^{j}]=\mathbb{E}[d\overline{W}_{t}^{j}dW_{t}^{i}]=\delta^{i;j}dt$.
In the steady state $\mathbb{E}_{\infty}[G_{ij}G_{ji}]=f(i,j)$ and $df(i,j)=0$
which leads to : 
\[
0=\underbrace{(\Delta_{i}^{\mathrm{dis}}+\Delta_{j}^{\mathrm{dis}})f(i,j)}_{\text{Laplacians}}+\underbrace{2(\delta_{i,j}-\delta_{i-1,j})n(i-1,i)+2(\delta_{i,j}-\delta_{i+1,j})n(i,i+1)}_{\text{Contact terms }}-\underbrace{\sum_{p\in\{0,L\}}\gamma_{p}(\delta_{i,p}+\delta_{j,p})f(i,j)}_{\text{Boundary}} .
\]
with $\gamma_k:=\alpha_k+\beta_k$ (i.e. $\gamma_0=1/a$ and $\gamma_L=1/b$). The terms weighted by the Kronecker delta's are what we will refer
to in the following as the {\it contact terms}. A stationarity equation is always composed of three parts as above :  the Laplacian terms, the contact
terms and the boundary terms.\par
We define the connected two-point expectations (for $i<j$):
\beq 
n^c(i,j):= [G_{ii}G_{jj}]-[G_{ii}][G_{jj}],\quad m^c(i):= [G_{ii}^2]-[G_{ii}]^2,\quad f(i,j)=[G_{ij}G_{ji}].
\eeq
The bulk stationarity equations impose that $(\Delta^\mathrm{dis}_i+\Delta^\mathrm{dis}_j)n^c(i,j)=0$ and $(\Delta^\mathrm{dis}_i+\Delta^\mathrm{dis}_j)f(i,j)=0$, with $\Delta^\mathrm{dis}_i$ is the discrete Laplacian as in the main text, which we enforce by demanding that 
\beq
 \Delta^\mathrm{dis}_i\,n^c(i,j)=\Delta^\mathrm{dis}_j\,n^c(i,j)=0,\quad \Delta^\mathrm{dis}_i\,f(i,j)=\Delta^\mathrm{dis}_j\,f(i,j)=0 .
 \eeq
The bulk/boundary stationarity equations demand that (with $j\not=0,L$ and $i\not=0,L$):
\begin{eqnarray}
n^c(1,j)-n^c(0,j) + \Delta^\mathrm{dis}_j n^c(0,j) &=& \gamma_a\, n^c(0,j),\\
f(1,j)-f(0,j) + \Delta^\mathrm{dis}_j f(0,j) &=& \gamma_a\, f(0,j),\\
n^c(i,L-1)-n^c(i,L) + \Delta^\mathrm{dis}_i n^c(i,L) &=& \gamma_b\, n^c(i,L),\\
f(i,L-1)-f(i,L) + \Delta^\mathrm{dis}_i f(i,L) &=& \gamma_b\, f(i,L),
\end{eqnarray}
The two first equations give $n^c(1,j)=(\gamma_a+1) n^c(0,j)$ and $f(1,j)=(\gamma_a+1) f(0,j)$. Using $\Delta^\mathrm{dis}_i\,n^c(i,j)=0$ and $\Delta^\mathrm{dis}_i\,f(i,j)=0$, this then recursively yields (for $i<j$):
\beq
 n^c(i,j)=(i \gamma_a+1) n^c(0,j),\quad f(i,j)=(i\gamma_a+1) f(0,j).
 \eeq
Similarly, starting from the other hand of the chain gives (for $i<j$):
\beq
 n^c(i,j)=( (L-j)\gamma_b+1) n^c(i,L),\quad f(i,j)=( (L-j)\gamma_b+1) f(i,L).
 \eeq
Hence, for $i<j$,
\beq 
n^c(i,j)=N\, (i+a)(L-j+b),\quad f(i,j)=F\, (i+a)(L-j+b),
\eeq
for some constants $N$ and $F$ (with $a=1/\gamma_a=1/(\alpha_0+\beta_0)$ and $b=1/\gamma_b=1/(\alpha_L+\beta_L)$.\\
The constants $N$ and $F$ are then determined by the stationarity conditions at the contact points $j=i$ and $j=i+1$. Stationarity of $m^c(i)$ yields
\beq \label{eq:m-discrete}
 2 m^c(i)= n^c(i,i+1)+n^c(i-1,i)+f(i,i+1)+f(i-1,i). 
 \eeq
The equation fixes $m^c(i)$ as a function of $f(i,j)$ and $n^c(i,j)$.
Using again the bulk relations $\Delta_i\,n^c(i,j)=0$ and $\Delta_i\,f(i,j)=0$, the stationarity conditions for $f(i,i+1)$ and for $n^c(i,i+1)$ give 
\begin{eqnarray}
m^c(i+1)+m^c(i) &=& 2 f(i,i+1) + n^c(i+1,i+1)+n^c(i,i),\\
m^c(i+1)+m^c(i)&=& 2n^c(i,i+1)+ f(i+1,i+1)+f(i,i)-\big(n^*(i+1)-n^*(i)\big)^2,
\end{eqnarray}
with $n^*(i)$ the mean profile.
Eliminating $m^c(i)$ from the above three equations give two equations for the constants $N$ and $F$:
\beq
 (F-N)(b+a+L-1)= (F+N)(a+b+L+1)=\big(n^*(i+1)-n^*(i)\big)^2.
 \eeq
Since $n^*(i+1)-n^*(i)=(\Delta n)/(b+a+L)$, with $\Delta n= n_b-n_a$, this yields
\begin{align}
 F &= \frac{(\Delta n)^2 }{(a+b+L-1)(a+b+L)(a+b+L+1)},\\
  N &= - \frac{(\Delta n)^2 }{(a+b+L-1)(a+b+L)^2(a+b+L+1)}.
 \end{align}
Hence
\begin{align}
  n^c(i,j) &=-\frac{(\Delta n)^2 (i+a)((L-j)+b)}{(a+b+L-1)(a+b+L)^2(a+b+L+1)},\\ 
   f(i,j) &= \frac{(\Delta n)^2 (i+a)((L-j)+b) }{(a+b+L-1)(a+b+L)(a+b+L+1)}.
  \end{align}
The cumulant $m^c(i)$ are then determined using (\ref{eq:m-discrete}):
\begin{align}
m^{c}(i) & =\frac{\Delta n^{2}\left(2(i+a)(L-i+b)-(L+b+a)\right)}{2(a+b+L)^{2}(a+b+L+1)}
\end{align}

{{\bf Cumulants of order 3}.--}  For $N=3$, with $0\leq i< j<k \leq L$, the non zero terms in the stationary state are : 
\begin{align}
[\hspace{4pt}\graphNcroixtrois{i}{j}{k}{0.5} \hspace{3pt} ]=[G_{ii}G_{jj}G_{kk}] &, \quad [\hspace{4pt} \graphNcroixFdeux{i}{j}{k}{0.5}]=[G_{ii}G_{jk}G_{kj}], \\
[\graphFdeuxcroixN{i}{j}{k}{0.5}\hspace{3pt}]=[G_{ij}G_{ji}G_{kk}] &, \quad  [\graphNcroixmFdeux{i}{j}{k}{0.5}]=[G_{ik}G_{jj}G_{ki}], \\
[\graphFtrois{i}{j}{k}{0.25}]=[G_{ij}G_{jk}G_{ki}] &, \quad [\graphNdeuxcroixN{i}{k}{0.5}\hspace{3pt}]=[G_{ii}^{2}G_{kk}], \\
[\hspace{4pt}\graphNcroixNdeux{i}{k}{0.5}]=[G_{ii}G_{kk}^{2}] &, \quad  [\graphNtrois{i}{0.5}]=[G_{ii}^{3}], \\
[\graphFdeuxpointN{i}{j}{0.7}]=[G_{ij}G_{ji}G_{jj}] &, \quad  [\graphNpointFdeux{i}{k}{0.7}]=[G_{ii}G_{ik}G_{ki}],
\end{align}
where the convention is that two indices written with different letters are evaluated at different sites. 
We define the connected expectations of the five first terms as :
\begin{align} \label{eq:diag-connexe}
[\hspace{0.17cm}\graphNcroixtrois{i}{j}{k}{0.5}\:]^{c} & =
[\hspace{0.17cm}\graphNcroixtrois{i}{j}{k}{0.5}\:] - [\hspace{0.17cm}\graphN{i}{0.5}\:][\hspace{0.17cm}\graphNcroixdeux{j}{k}{0.5}\:]-
[\hspace{0.17cm}\graphN{j}{0.5}\:][\hspace{0.17cm}\graphNcroixdeux{k}{i}{0.5}\:] - [\hspace{0.17cm}\graphNcroixdeux{i}{j}{0.5}\:][\hspace{0.17cm}\graphN{k}{0.5}\:]
+2[\hspace{0.17cm}\graphN{i}{0.5}\:][\hspace{0.17cm}\graphN{j}{0.5}\:][\hspace{0.17cm}\graphN{k}{0.5}\:] \\
[\hspace{0.17cm}\graphNcroixFdeux{i}{j}{k}{0.5}]^{c} & =[\hspace{0.17cm}\graphNcroixFdeux{i}{j}{k}{0.5}]-[\hspace{0.17cm}\graphN{i}{0.5}\:][\graphFdeux{j}{k}{0.5}]  \\{}
[\graphFdeuxcroixN{i}{j}{k}{0.5}\:]^{c} & =[\graphFdeuxcroixN{i}{j}{k}{0.5}\:]-[\graphFdeux{i}{j}{0.5}][\hspace{0.17cm}\graphN{k}{0.5}\:]\\{}
[\graphNcroixmFdeux{i}{j}{k}{0.5}]^{c} & =[\graphNcroixmFdeux{i}{j}{k}{0.5}]-[\graphFdeux{i}{k}{0.5}][\hspace{0.17cm}\graphN{j}{0.5}\:]  \\{}
[\graphFtrois{i}{j}{k}{0.25}]^{c} & =[\graphFtrois{i}{j}{k}{0.25}]\ 
\end{align}

We name these terms $g^{c}_m(i,j,k)$ where the index $m$ runs from
$1$ to $5$ and labels the diagrams. The stationarity conditions for the $g^{c}_m$ follow all the same
pattern :
\begin{align}
0 & =(\Delta_{i}^{\mathrm{dis}}+\Delta_{j}^{\mathrm{dis}}+\Delta_{k}^{\mathrm{dis}})(g^{c}_m(i,j,k))+\text{``contact terms''}\\
 & \hspace{1em}-(\gamma_{0}\delta_{i,0}+\gamma_{L}\delta_{k,L})g^{c}_m(i,j,k) \nonumber
\end{align}
where the contact terms are non zero only when two indices are close
to each other : $j=i+1$, $k=j+1$ and depends on $m$. As with the
$N=2$ case, using bulk and boundary stationarity, we find that all
the $g^{c}_m$'s must be of the following polynomial form : 
\[
g^{c}_m(i,j,k)=N_{m}(a+i)(1+C_{m}j)(L+b-k)
\]
We make use of the contact terms to determine the constant $N_{m}$
and $C_{m}$. We will not show the explicit derivation as it does
not entail any difficulties but is a bit cumbersome. Knowing the $g^{c}_m$'s,
it is then easy to retrieve the remaining terms by again making use
of the various contact terms. The result is :
\begin{flalign}
[\hspace{0.17cm}\graphNcroixtrois{i}{j}{k}{0.5}\:]^{c} & =-\frac{4(\Delta n)^{3}(a+i)(a-b+2j-L)(L+b-k)}{(L+a+b-2)(L+a+b-1)(L+a+b)^{3}(L+a+b+1)(L+a+b+2)}\\{}
[\hspace{0.17cm}\graphNcroixFdeux{i}{j}{k}{0.5}]^{c} & =\frac{2(\Delta n)^{3}(a+i)(a-b+2j-L)(L+b-k)}{(L+a+b-2)(L+a+b-1)(L+a+b)^{2}(L+a+b+1)(L+a+b+2)}\\{}
[\graphFdeuxcroixN{i}{j}{k}{0.5}\:]^{c} & =[\graphNcroixmFdeux{i}{j}{k}{0.5}]^{c}=[\hspace{0.17cm}\graphNcroixFdeux{i}{j}{k}{0.5}]^{c}\\{}
[\graphFtrois{i}{j}{k}{0.25}]^{c} & =-\frac{(\Delta n)^{3}(a+i)(-L-b+a+2j)(L+b-k)}{(L+a+b-2)(L+a+b-1)(L+a+b)(L+a+b+1)(L+a+b+2)}\\{}
[\:\graphNdeuxcroixN{i}{j}{0.5}\:]^{c} & =\frac{(\Delta n)^{3}\left(2(a+i)(-L-b+a+2i)+L+a+b\right)(L+b-j)}{(L+a+b-1)(L+a+b)^{3}(L+a+b+1)(L+a+b+2)}\\{}
[\hspace{0.17cm}\graphNcroixNdeux{i}{j}{0.5}\:]^{c} & =\frac{(\Delta n)^{3}(a+i)(a(2(L+b-j)-1)-2(L+b-2j)(L+b-j)-L-b)}{(L+a+b-1)(L+a+b)^{3}(L+a+b+1)(L+a+b+2)}\\{}
[\graphNpointFdeux{i}{j}{0.7}]^{c} & =-\frac{(\Delta n)^{3}\left(2(a+i)(-L-b+a+2i)+L+a+b\right)(L+b-j)}{2(L+a+b-1)(L+a+b)^{2}(L+a+b+1)(L+a+b+2)}\\{}
[\graphFdeuxpointN{i}{j}{0.7}]^{c} & =-\frac{(\Delta n)^{3}(a+i)(a(2(L+b-j)-1)-2(L+b-2j)(L+b-j)-L-b)}{2(L+a+b-1)(L+a+b)^{2}(L+a+b+1)(L+a+b+2)}\\{}
[\graphNtrois{i}{0.5}]^{c} & =-\frac{(\Delta n)^{3}(L+a+b-2)(-L-b+a+2i)(a(4(L+b-i)-3)+4i(L+b-i)-3(b+L))}{4(L+a+b-1)(L+a+b)^{3}(L+a+b+1)(L+a+b+2)}
\end{flalign}
where the connected expectations are defined according to eqs.(\ref{eq:diag-connexe}) with appropriate identification of the indices.
For $N=3$ the leading order in the thermodynamic limit is $O(L^{-2})$.
These include $[\graphNpointFdeux{i}{j}{0.7}]^{c}$, $[\graphFdeuxpointN{i}{j}{0.7}]^{c}$,
$[\graphNtrois{i}{0.5}]^{c}$ and $[\graphFtrois{i}{j}{k}{0.25}]^{c}$.
In the thermodynamic limit, with $x= i/L$, $y= j/L$, $z= k/L$,
they read : 
\begin{align}
[\graphNpointFdeux{i}{j}{0.7}]^{c} & =\frac{(\Delta n)^{3}x(1-2x)(1-y)}{L^{2}}\\{}
[\graphFdeuxpointN{i}{j}{0.7}]^{c} & =\frac{(\Delta n)^{3}x(1-2y)(1-y)}{L^{2}}\\{}
[\graphNtrois{i}{0.5}]^{c} & =\frac{(\Delta n)^{3}x(1-2x)(1-x)}{L^{2}}\\{}
[\graphFtrois{i}{j}{k}{0.25}] & =\frac{(\Delta n)^{3}x(1-2y)(1-z)}{L^{2}}
\end{align}
As stated in the main text we see that in the thermodynamic limit,
the knowledge of the single loop $[\graphFtrois{i}{j}{k}{0.25}]$
is enough to get the other terms by continuity.  

\section{Conditions for stationarity, blow-ups and higher order cumulants}

We recall the stochastic equations (\ref{eq:Main_sto1},\ref{eq:Main_sto2},\ref{eq:Main_sto3}).
\medskip

{{\bf Conditions for stationarity and blow-ups.}--}
We look at expectations of the following form
\beq
\bigg[\singleloop{i}{j}{0.5}\bigg] , \quad
\bigg[\separes{i}{j}{0.5}\bigg] , \quad
\bigg[\contact{i}{0.5}\bigg] ,   
\eeq
where $A$ and $B$ are subgraphs (We could also add another subgraph linking $A$ and $B$).

We write the stationarity conditions, using (\ref{eq:Main_sto1},\ref{eq:Main_sto2},\ref{eq:Main_sto3}). The first set of bulk relations, for $i$ and $j$ far a part and away from the boundaries, are satisfied if
\beq \label{eq:Laplace} 
\Delta^\mathrm{dis}_i\, \bigg[\singleloop{i}{j}{0.5}\bigg] =\Delta^\mathrm{dis}_j\, \bigg[\singleloop{i}{j}{0.5}\bigg]=0,\quad
\Delta^\mathrm{dis}_i\, \bigg[\separes{i}{j}{0.5}\bigg]=\Delta^\mathrm{dis}_j\, \bigg[\separes{i}{j}{0.5}\bigg] =0,
\eeq
with $\Delta^\mathrm{dis}_i$ the discrete Laplacian as in the main text.
Notice that $\bigg[ \contact{i}{0.5}\bigg]$ is not discrete harmonic. 

As for the two- and three- point functions, it is easy to see that the stationarity conditions imposes the connected components of $\bigg[\singleloop{i}{j}{0.5}\bigg] $ and $\bigg[ \separes{i}{j}{0.5}\bigg]$ to vanish at the boundaries in the large size limit.

We then write the contact conditions which come from the stationarity conditions for $j=i$ or $j=i\pm 1$. We start with $j=i\pm 1$. The case $j=i-1$ is recovered from the case $j=i+1$ up to the exchange of $A$ and $B$. Using (\ref{eq:Laplace}), stationarity of $\bigg[\singleloop{j}{j+1}{0.5}\bigg] $ gives
\beq \label{eq:Stat1(j;j+1)}
\bigg[\contact{j}{0.5}\bigg]
+ \bigg[\contact{j+1}{0.5}\bigg]
= \bigg[\singleloop{j}{j}{0.5}\bigg] 
+ \bigg[\singleloop{j+1}{j+1}{0.5}\bigg]
+ \bigg[\separes{j}{j+1}{0.5}\bigg]
+ \bigg[\separes{j+1}{j}{0.5}\bigg]
\eeq
Similarly, stationarity of $ \bigg[\separes{j}{j+1}{0.5}\bigg]$ gives
\beq \label{eq:Stat2(j;j+1)}
\bigg[\contact{j}{0.5}\bigg]
+ \bigg[\contact{j+1}{0.5}\bigg]
= \bigg[\separes{j}{j}{0.5}\bigg]+ \bigg[\separes{j+1}{j+1}{0.5}\bigg]
+ \bigg[\singleloop{j+1}{j}{0.5}\bigg]  
+ \bigg[\singleloop{j}{j+1}{0.5}\bigg] 
\eeq
The stationarity conditions for $\bigg[\contact{j}{0.5}\bigg]$ yields
\begin{align} \label{eq:Stat1(j;j)}
4\, \bigg[\contact{j}{0.5}\bigg]
&= \bigg[\singleloop{j+1}{j}{0.5}\bigg]+ \bigg[\singleloop{j}{j+1}{0.5}\bigg]
+ \bigg[\singleloop{j-1}{j}{0.5}\bigg]+\bigg[\singleloop{j}{j-1}{0.5}\bigg] \\
&+ \bigg[\separes{j+1}{j}{0.5}\bigg]+\bigg[\separes{j}{j+1}{0.5}\bigg]
+ \bigg[\separes{j-1}{j}{0.5}\bigg]+ \bigg[\separes{j}{j-1}{0.5}\bigg]
\nonumber
\end{align}

\bigskip 

{{\bf Stationarity conditions  for higher order cumulants.}--}
Eliminating $\bigg[\contact{j}{0.5}\bigg]
+ \bigg[\contact{j+1}{0.5}\bigg]$ from the two equations (\ref{eq:Stat1(j;j+1)},\ref{eq:Stat2(j;j+1)}) yields
\begin{align} \label{eq:StoDiff}
& \bigg[\singleloop{j+1}{j+1}{0.5}\bigg]- \bigg[\singleloop{j}{j+1}{0.5}\bigg]
+ \bigg[\singleloop{j}{j}{0.5}\bigg]- \bigg[\singleloop{j+1}{j}{0.5}\bigg] \\
&=\bigg[\separes{j+1}{j+1}{0.5}\bigg]- \bigg[\separes{j}{j+1}{0.5}\bigg]
+ \bigg[\separes{j}{j}{0.5}\bigg]- \bigg[\separes{j+1}{j}{0.5}\bigg]
\nonumber
\end{align}
The l.h.s. is the difference of discrete derivatives. In the large size limit,  the r.h.s. is dominated by the disconnected contributions to the expectations because the approximation 
\beq 
\bigg[\separes{i}{j}{0.5}\bigg] = 
\bigg[\oneloopg{i}{0.5}\bigg] \bigg[\oneloopd{j}{0.5}\bigg]+ \cdots,
\eeq
where the dots refer to sub-leading terms in $1/L$, is valid in the large size limit. Hence, equation (\ref{eq:StoDiff}) can be written as
\beq
\nabla^\mathrm{dis}_x\, \bigg[\singleloop{x}{y}{0.5}\bigg]\big\vert_{y=x^+}
- \nabla^\mathrm{dis}_y \, \bigg[\singleloop{y}{x}{0.5}\bigg]\big\vert_{y=x^+}
= \bigg({\nabla^\mathrm{dis}_x \bigg[\oneloopg{x}{0.5}\bigg] }\bigg) \cdot \left(\nabla^\mathrm{dis}_y \bigg[\oneloopd{y}{0.5}\bigg]\right)\bigg\vert_{y=x}
\eeq
up to sub-leading terms in $1/L$, where we adopt a more explicit continuous indexation (i.e. $x=i/L$, etc.), and with $\nabla^\mathrm{dis}_xf(x)= f(x+1/L)-f(x)\simeq L^{-1} \nabla_xf(x)$ in the large $L$ limit.

Exchanging the role of $A$ and $B$ (which amounts to exchange $x$ and $y$) gives two relations. Exchanging and taking the sum yields (in the large $L$ limit)
\beq \label{eq:final-discrete}
\left({ \nabla^\mathrm{dis}_x- \nabla^\mathrm{dis}_y }\right)\, \left(\bigg[\singleloop{x}{y}{0.5}\bigg]+\bigg[\singleloop{y}{x}{0.5}\bigg]\right)\bigg\vert_{y=x^+}
=2 \bigg({\nabla^\mathrm{dis}_x \bigg[\oneloopg{x}{0.5}\bigg] }\bigg) \cdot \left(\nabla^\mathrm{dis}_y \bigg[\oneloopd{y}{0.5}\bigg]\right)\bigg\vert_{y=x}
\eeq
Exchanging and taking the difference yields (in the large $L$ limit)
\beq \label{eq:continu-discrete}
\left({ \nabla^\mathrm{dis}_x + \nabla^\mathrm{dis}_y }\right)\, \left(\bigg[\singleloop{x}{y}{0.5}\bigg]-\bigg[\singleloop{y}{x}{0.5}\bigg]\right)\bigg\vert_{y=x^+}
= 0.
\eeq

If $\bigg[\singleloop{}{}{0.5}\bigg]$ is a single loop diagram, then $A$ and $B$  are circle arcs and $\bigg[\oneloopg{}{0.5}\bigg] $ and $\bigg[\oneloopd{}{0.5}\bigg]$ are single loops with respectively $N_a$ and $N_b$ edges (so that the parent single loop has $N=N_a+N_b$ edges). By recursion, the expectations $\bigg[\oneloopg{}{0.5}\bigg] $ and $\bigg[\oneloopd{}{0.5}\bigg]$ scale respectively as $1/L^{N_a-1}$ and $1/L^{N_b-1}$, and the l.h.s. of (\ref{eq:final-discrete}) scales as $1/L^{N_a-1+N_b-1+2}=1/L^N$. Hence, by (\ref{eq:final-discrete}), $\bigg[\singleloop{}{}{0.5}\bigg]$ scales as $1/L^{N-1}$. Equation (\ref{eq:final-discrete}) then reads
\beq \label{eq:final}
\left({ \nabla_x- \nabla_y }\right)\, \left({ 
\bigg[\singleloop{x}{y}{0.5}\bigg]
+ \bigg[\singleloop{y}{x}{0.5}\bigg] }\right)\bigg\vert_{y=x^+}
= \frac{2}{L}\, \left({\nabla_x \bigg[\oneloopg{x}{0.5}\bigg] }\right) \cdot \left({\nabla_y \bigg[\oneloopd{y}{0.5}\bigg]  }\right)\bigg\vert_{y=x}
\eeq
which coincides with the main contact relation of the main text.

Similarly, in the large size limit, equation (\ref{eq:continu-discrete}) becomes the continuity equation
\beq \label{eq:continu-final}
\bigg[\singleloop{x}{y}{0.5}\bigg]\bigg\vert_{y=x^+}
= \bigg[\singleloop{y}{x}{0.5}\bigg]\bigg\vert_{y=x^+}
\eeq

Finally, equation (\ref{eq:Stat1(j;j)}) in the large size limit gives
\beq
\bigg[\contact{x}{0.5}\bigg]^c 
= \bigg[\singleloop{x}{x}{0.5}\bigg] ,
\eeq
which says that connected expectations of pinched diagrammes are obtained by continuity from the parent diagramme.
Furthermore, we can use the stationarity equations (\ref{eq:Stat1(j;j+1)},\ref{eq:Stat2(j;j+1)},\ref{eq:Stat1(j;j)}) to prove that $\bigg[\separes{x}{y}{0.5}\bigg] ^c$ is sub-leading compare to $ \bigg[\singleloop{x}{y}{0.5}\bigg]$ by a factor $1/L$.

\bigskip 

{{\bf Explicit solutions for the first cases.}--}
Here, we solve the stationarity conditions for the first cumulants. Recall that $n^*(x)=n_a+ x(n_b-n_a)$. The case $N=2$ was done in the main text (from the discrete solution) with output
\beq 
\Big[\graphFdeux{x\hspace{0.4 cm}}{\hspace{0.4 cm}y}{0.5}\Big] = \frac{1}{L}\, f_2(x,y),\quad f_2(x,y)= (\Delta n)^2\, x(1-y).
\eeq
For $N=3$, there are only one diagram
\beq
\Big[\graphFtrois{x}{y}{z}{0.3}\Big] = \frac{1}{L^2}\, f_3(x,y,z),
\eeq
with $0\leq x<y<z\leq1$ by convention. Because there is only one sector in the case $N=3$, the equations for $f_3(x,y,z)$ simplify to
\begin{gather}
\Delta_{x}\bigg[\graphFtrois{x}{y}{z}{0.3}\bigg]=\Delta_{y}\bigg[\graphFtrois{x}{y}{z}{0.3}\bigg]=\Delta_{z}\bigg[\graphFtrois{x}{y}{z}{0.3}\bigg]=0\\
\bigg[\graphFtrois{0}{y}{z}{0.3}\bigg]=\bigg[\graphFtrois{x}{y}{1}{0.3}\bigg]=0\\
(\nabla_{x}-\nabla_{y})\bigg(\bigg[\graphFtrois{x}{y}{z}{0.3}\bigg]\bigg)\bigg\vert_{x=y}=\frac{1}{L}\big(\nabla_{x}\Big[\hspace{0.3cm}\graphN{x}{0.6}\:\Big]\big)\cdot \big(\nabla_{y}\Big[\graphFdeux{y\hspace{1em}}{\hspace{1em}z}{0.5}\Big]\big)\big\vert_{x=y}\\
(\nabla_{y}-\nabla_{z})\bigg(\bigg[\graphFtrois{x}{y}{z}{0.3}\bigg]\bigg)\bigg\vert_{y=z}=\frac{1}{L}\big(\nabla_{y}\Big[\hspace{0.3cm}\graphN{y}{0.6}\:\Big]\big)\cdot \big(\nabla_{z}\Big[\graphFdeux{x\hspace{1em}}{\hspace{1em}z}{0.5}\Big]\big)\big\vert_{y=z}
\end{gather}
The solution is of the form $f_3(x,y,z)=x\,Q(y)\,(1-z)$ for some polynomial $Q(y)$ of degree one. Solving for it using the above equations gives
\[\bigg[ \graphFtrois{x}{y}{z}{0.3}\bigg]=\frac{(\Delta n)^3}{L^2}x(1-2y)(1-z) .\]
For $N=4$ there are a priori 3 different types of one-loop diagrams : 
$\bigg[\graphFquatre{x_{1}}{x_{2}}{x_{3}}{x_{4}}{0.5}\bigg]$, $\bigg[\graphFquatre{x_{2}}{x_{1}}{x_{3}}{x_{4}}{0.5}\bigg]$,
and $\bigg[\graphFquatre{x_{1}}{x_{3}}{x_{2}}{x_{4}}{0.5}\bigg]$
with $0\leq x_{1}<x_{2}<x_{3}<x_{4}\leq1$. They fulfill the following
equations : 
\begin{gather}
\Delta_{x_{j}}\bigg[\graphFquatre{x_{1}}{x_{2}}{x_{3}}{x_{4}}{0.5}\bigg]=\Delta_{x_{j}}\bigg[\graphFquatre{x_{2}}{x_{1}}{x_{3}}{x_{4}}{0.5}\bigg]=\Delta_{x_{j}}\bigg[\graphFquatre{x_{1}}{x_{3}}{x_{2}}{x_{4}}{0.5}\bigg]=0,\hspace{1em}\forall j\in\llbracket1,4\rrbracket\\
\bigg[\graphFquatre{0}{x_{2}}{x_{3}}{x_{4}}{0.5}\bigg]=\bigg[\graphFquatre{x_{2}}{0}{x_{3}}{x_{4}}{0.5}\bigg]=\bigg[\graphFquatre{0}{x_{3}}{x_{2}}{x_{4}}{0.5}\bigg]=0\\
\bigg[\graphFquatre{x_{1}}{x_{2}}{x_{3}}{1}{0.5}\bigg]=\bigg[\graphFquatre{x_{2}}{x_{1}}{x_{3}}{1}{0.5}\bigg]=\bigg[\graphFquatre{x_{1}}{x_{3}}{x_{2}}{1}{0.5}\bigg]=0\\
(\nabla_{x_{1}}-\nabla_{x_{2}})\bigg(\bigg[\graphFquatre{x_{1}}{x_{2}}{x_{3}}{x_{4}}{0.5}+\graphFquatre{x_{2}}{x_{1}}{x_{3}}{x_{4}}{0.5}\bigg]\bigg)\bigg\vert_{x_{1}=x_{2}}=\frac{2}{L}\big(\nabla_{x_{1}}\Big[\hspace{0.5cm}\graphN{x_{1}}{0.6}\:\Big]\big)\cdot \big(\nabla_{x_{2}}\Big[\graphFtrois{x_{4}}{x_{2}}{x_{3}}{0.3}\Big]\big)\big\vert_{x_{1}=x_{2}}\\
\lim_{x_{1}\to x_{2}}\bigg(\bigg[\graphFquatre{x_{1}}{x_{2}}{x_{3}}{x_{4}}{0.5}\bigg]-\bigg[\graphFquatre{x_{2}}{x_{1}}{x_{3}}{x_{4}}{0.5}\bigg]\bigg)=0\\
(\nabla_{x_{2}}-\nabla_{x_{3}})\bigg(\bigg[\graphFquatre{x_{1}}{x_{2}}{x_{3}}{x_{4}}{0.5}+\graphFquatre{x_{1}}{x_{3}}{x_{2}}{x_{4}}{0.5}\bigg]\bigg)\bigg\vert_{x_{2}=x_{3}}=\frac{2}{L}\big(\nabla_{x_{2}}\Big[\hspace{0.5cm}\graphN{x_{2}}{0.6}\:\big]\big)\cdot \big(\nabla_{x_{3}}\Big[\graphFtrois{x_{1}}{x_{3}}{x_{4}}{0.3}\Big]\big)\big\vert_{x_{2}=x_{3}}\\
\lim_{x_{2}\to x_{3}}\bigg(\bigg[\graphFquatre{x_{1}}{x_{2}}{x_{3}}{x_{4}}{0.5}\bigg]-\bigg[\graphFquatre{x_{1}}{x_{3}}{x_{2}}{x_{4}}{0.5}\bigg]\bigg)=0\\
(\nabla_{x_{3}}-\nabla_{x_{4}})\bigg(\bigg[\graphFquatre{x_{1}}{x_{2}}{x_{3}}{x_{4}}{0.5}+\graphFquatre{x_{2}}{x_{1}}{x_{3}}{x_{4}}{0.5}\bigg]\bigg)\bigg\vert_{x_{3}=x_{4}}=\frac{2}{L}\big(\nabla_{x_{3}}\Big[\graphFtrois{x_{1}}{x_{2}}{x_{3}}{0.3}\Big]\big)\cdot \big(\nabla_{x_{4}}\Big[\hspace{0.5cm}\graphN{x_{4}}{0.6}\:\Big]\big)\big\vert_{x_{3}=x_{4}}\\
\lim_{x_{3}\to x_{4}}\bigg(\bigg[\graphFquatre{x_{1}}{x_{2}}{x_{3}}{x_{4}}{0.5}\bigg]-\bigg[\graphFquatre{x_{2}}{x_{1}}{x_{3}}{x_{4}}{0.5}\bigg]\bigg)=0\\
(\nabla_{x_{2}}-\nabla_{x_{3}})\bigg(\bigg[\graphFquatre{x_{2}}{x_{1}}{x_{3}}{x_{4}}{0.5}\bigg]\bigg)\bigg\vert_{x_{2}=x_{3}}=\frac{1}{L}\nabla_{x_{2}}(\big[\graphFdeux{x_{1}\hspace{1em}}{\hspace{1em}x_{2}}{0.5}\big])\cdot \nabla_{x_{3}}(\big[\graphFdeux{x_{3}\hspace{1em}}{\hspace{1em}x_{4}}{0.5}\big])\big\vert_{x_{2}=x_{3}}\\
(\nabla_{x_{1}}-\nabla_{x_{2}})\bigg(\bigg[\graphFquatre{x_{1}}{x_{3}}{x_{2}}{x_{4}}{0.5}\bigg]\bigg)\bigg\vert_{x_{1}=x_{2}}=\frac{1}{L}\nabla_{x_{1}}(\big[\graphFdeux{x_{1}\hspace{1em}}{\hspace{1em}x_{3}}{0.5}\big])\cdot \nabla_{x_{2}}(\big[\graphFdeux{x_{2}\hspace{1em}}{\hspace{1em}x_{4}}{0.5}\big])\big\vert_{x_{1}=x_{2}}\\
(\nabla_{x_{3}}-\nabla_{x_{4}})\bigg(\bigg[\graphFquatre{x_{1}}{x_{3}}{x_{2}}{x_{4}}{0.5}\bigg]\bigg)\bigg\vert_{x_{3}=x_{4}}=\frac{1}{L}\nabla_{x_{3}}(\big[\graphFdeux{x_{1}\hspace{1em}}{\hspace{1em}x_{3}}{0.5}\big])\cdot \nabla_{x_{4}}(\big[\graphFdeux{x_{2}\hspace{1em}}{\hspace{1em}x_{4}}{0.5}\big])\big\vert_{x_{3}=x_{4}}
\end{gather}
The bulk/boundary conditions impose that all loops must be polynomial
of the form $x_{1}Q(x_{2},x_{3})(1-x_{4})$. Using the remaining conditions,
one gets :
\begin{align}
\bigg[\graphFquatre{x_{1}}{x_{2}}{x_{3}}{x_{4}}{0.5}\bigg] & =\frac{(\Delta n)^{4}}{L^{3}}x_{1}(1-3x_{2}-2x_{3}+5x_{2}x_{3})(1-x_{4})\\
\bigg[\graphFquatre{x_{2}}{x_{1}}{x_{3}}{x_{4}}{0.5}\bigg] & =\frac{(\Delta n)^{4}}{L^{3}}x_{1}(1-3x_{2}-2x_{3}+5x_{2}x_{3})(1-x_{4})\\
\bigg[\graphFquatre{x_{1}}{x_{3}}{x_{2}}{x_{4}}{0.5}\bigg] & =\frac{(\Delta n)^{4}}{L^{3}}x_{1}(1-4x_{2}-x_{3}+5x_{2}x_{3})(1-x_{4})
\end{align}

\section{Classical SSEP}

Let us first recall the connection between the quantum and classical SSEP models:
\beq \label{eq:class-SSEP}
 \langle e^{\sum_j a_j \mathfrak{n}_j}\rangle_\mathrm{ssep} = \mathrm{Tr}\big( \bar \rho\, e^{\sum_i a_i \hat n_i}\big)
= \Big[ \mathrm{Tr}\big( \rho\, e^{\sum_i a_i \hat n_i}\big)\Big] ,
\eeq
with $\hat n_i=c^\dag_ic_i$ the quantum number operators and $\mathfrak{n}_j$ the classical SSEP occupation variables.

In this part, we prove eq.(17) of the main text as a consequence of eq.(8) in the main text. First, using Wick's theorem we have  :
\beq \label{eq:bigN}
\Big[\Tr(\rho\, \hat{n}_{i_{1}}\hat{n}_{i_{2}}\cdots\hat{n}_{i_{N}})\Big]=\sum_{{\cal P}=\{u_{1},\cdots,u_{m}\}}\sum_{\{\sigma_{u_{1}}\cdots\sigma_{u_{m}}\}}(-1)^{N+m}\bigg[\graphFquatre{i_{\sigma_{u_{1}}(1)}}{i_{\sigma_{u_{1}}(2)}}{\ldots}{i_{\sigma_{u_{1}}(u_{1})}}{0.5}\ldots\graphFquatre{i_{\sigma_{u_{m}}(N-u_{m}+1)}}{i_{\sigma_{u_{m}}(N-u_{m}+1)}}{\ldots}{i_{\sigma_{u_{m}}(N)}}{0.5}\bigg]
\eeq
where ${\cal P}=\{u_{1},\cdots,u_{m}\}$ are partitions of $N$, i.e. $\sum_{j=1}^{m}u_{j}=N$ and the $\sigma_{p}$'s denote all permutations of $p$ indices up to cyclic permutations (there are $(p-1)!$ of them). This formula is for $i_{1}\neq i_{2}\neq\cdots\neq i_{N}$, and taking two indices to be equal amounts to consider a term of order $N-1$ since $\hat{n}^{2}=\hat{n}$.
The graph $\graphFquatre{i_{1}}{i_{2}}{\ldots}{i_{p}}{0.5}$ designates a one-loop diagram connecting $p$ points. 
One can directly check that $\Big[\Tr(\rho\, \hat{n}_{i_{1}}\hat{n}_{i_{2}}\cdots\hat{n}_{i_{N}})\Big]$ equals the terms of order $a_{i_{1}}\cdots a_{i_{N}}$, in 
\begin{equation} \label{expaini}
\bigg[\exp\big(\sum_{p=1}^{\infty}\sum_{i_{1}\cdots i_{p}}\frac{(-1)^{p-1}}{p!}a_{i_{1}}\cdots a_{i_{p}}\,\sum_{\sigma_{p}}\graphFquatre{i_{\sigma_{p}(1)}}{i_{\sigma_{p}(2)}}{\ldots}{i_{\sigma_{p}(p)}}{0.5} \big) \bigg] . 
\end{equation}

The two last formula can also be retrieved from the well known formula for fermionic expectation values:
\beq
\Tr\big(\rho\,\exp(\sum_{i}a_{i}\hat{n}_{i})\big) = \mathrm{Det}\big( 1 + G(e^A-1)\big).
\eeq
with $A$ the diagonal matrix with entries $a_i$.
Expanding the determinant as a sum over permutations and decomposing these permutations into product of cycles yields (\ref{eq:bigN}).

Recall the definition of the connected correlation function $[X_{1},\cdots,X_{q}]^{c}$ of random variables $X_k$:
\[
[X_{1},\cdots,X_{q}]^{c}=\frac{\partial}{\partial a_{1}}...\frac{\partial}{\partial a_{q}}\log\big( \exp(\sum_{i=1}^{q}a_{i}X_{i})\big)\,\vert_{a_{1}=a_{2}...=a_{q}=0}
\]
Combining this definition with (\ref{expaini}) and the correspondance between the quantum and classical SSEP correlations, it is clear the the classical SSEP expectations are given by : 
\begin{align}
\langle \mathfrak{n}_{i_1}\cdots  \mathfrak{n}_{i_N} \rangle^c_\mathrm{ssep}
=\sum_{{\cal P}=\{u_{1},\cdots,u_{m}\}}\sum_{\{\sigma_{u_{1}}\cdots\sigma_{u_{m}}\}}(-1)^{N+m}\bigg[\graphFquatre{i_{\sigma_{u_{1}}(1)}}{i_{\sigma_{u_{1}}(2)}}{\ldots}{i_{\sigma_{u_{1}}(u_{1})}}{0.5}\ldots\graphFquatre{i_{\sigma_{u_{m}}(N-u_{m}+1)}}{i_{\sigma_{u_{m}}(N-u_{m}+1)}}{\ldots}{i_{\sigma_{u_{m}}(N)}}{0.5}\bigg]^{c} ,
\end{align}
 for all $i_k$ distinct. The only remaining terms in the thermodynamic limit among the connected diagrams are the one-loop diagrams, thus
proving eq.(17) of the main text : 
\beq 
\langle\mathfrak{n}_{j_{1}}...\mathfrak{n}_{j_{N}}\rangle_{\mathrm{ssep}}^{c}=\frac{(-)^{N-1}}{L^{N-1}}\sum_{\sigma}f_{N}^{\sigma}({\bm{x}})+O(L^{-N}).
\eeq
Up to order four, we checked that this formula indeed agreed with known results for SSEP (see references [6,7,76] of the main text):
\begin{align*}
\langle\mathfrak{n}_{j_{1}}\rangle_{\mathrm{ssep}} & =[\hspace{0.5cm}\graphN{x_{1}}{0.6}\:]=n_{a}+(\Delta n)\,x_{1}\\
\langle\mathfrak{n}_{j_{1}}\mathfrak{n}_{j_{2}}\rangle_{\mathrm{ssep}}^{c} & =- \big[\graphFdeux{x_{1}\hspace{1em}}{\hspace{1em}x_{3}}{0.5}\big]=\frac{(\Delta n)^{2}}{L}x_{1}(1-x_{2})\\
\langle\mathfrak{n}_{j_{1}}\mathfrak{n}_{j_{2}}\mathfrak{n}_{j_{3}}\rangle_{\mathrm{ssep}}^{c} & = 2 \bigg[\graphFtrois{x_{1}}{x_{2}}{x_{3}}{0.3}\bigg]\\
 & =\frac{(\Delta n)^{3}}{L^{2}}2x_{1}(1-2x_{2})(1-x_{3})\\
\langle\mathfrak{n}_{j_{1}}\mathfrak{n}_{j_{2}}\mathfrak{n}_{j_{3}}\mathfrak{n}_{j_{4}}\rangle_{\mathrm{ssep}}^{c} & =- 2 \bigg[\graphFquatre{x_{1}}{x_{2}}{x_{3}}{x_{4}}{0.5}+\graphFquatre{x_{2}}{x_{1}}{x_{3}}{x_{4}}{0.5}+\graphFquatre{x_{1}}{x_{3}}{x_{2}}{x_{4}}{0.5}\bigg]\\
 & =-\frac{(\Delta n)^{4}}{L^{3}}2x_{1}(3-10x_{2}-5x_{3}+15x_{2}x_{3})(1-x_{4})
\end{align*}
Notice again that the classical SSEP cumulants are given by the sum over the different sectors of the quantum SSEP single loops.

\section{Discussion}

This  set of remarks aims at making clear within which context our work has to be perceived.  They arose while  answering  the  following  questions  asked  by  the referees: What is the simplest non-classical observable that an experiment (or a numerical simulation) should look for, and how different is its behaviour from what would have been expected in a purely classical model? What kind of physics is to be elucidated through the study such kind of model? How general is the method of solution, and is there any connection between the notion of integrability here and the standard notion of quantum integrability via mapping to a Hubbard model in [69]?

The short answer to the first question is ``look for coherent effects", in the form of say interferences or entanglements. Indeed, as explained in the text and noticed by the referees, the correlation functions of local density observables are in one-to-one correspondence with their classical counterparts. To see echoes of quantum fluctuations, one should then look after quantum coherent effects, and we have shown that these quantum coherences (encoded in the off-diagonal elements of the matrix of two-point functions) scale as the inverse of the square root of the system size (to leading order). 
One, possibly simple, experimental route towards this would be to conduct an interferometry experiment between two parts of the system. Such setup has recently been discussed by M.J. Gullans and D.A. Huse in [57] and their proposal can be directly transposed to our system. It consists in connecting two distant sites of the lattice to tunnelling barriers into 1D channels. The tunnelling fermions are then used for an interferometric experiment via, for instance, a beam splitter. The quantum fluctuating coherences will be visible in the interference patterns.
Another possibility consists in imposing different boundary conditions. The boundary conditions considered in the text amount to injecting and extracting fermions in the system at given rates, in occupancy eigenstates. Since our model is quantum mechanical, one could instead consider injecting/extracting fermions in states that are linear superpositions of occupied and empty sites. These boundary conditions then impose new steady distributions for which there is not an one-to-one correspondence between the quantum occupancy operator correlations and classical ones anymore. One should then be able to observe the difference between the classical and quantum systems by simply measuring the density profiles of the system. This new setting however deserves a study of its own (to which we hope/plan to come back soon) and we prefer not discuss it here (by lack of room).

As explained in the introduction, the kind of physics we aim at deciphering with this model concerns those covered by the extension of the macroscopic fluctuation theory, which is a classical theory, to the quantum realm. The macroscopic fluctuation theory is a quite general framework adapted to describe properties of locally diffusive classical systems away from equilibrium. Its quantum version would hopefully have a large degree of generality and apply for quantum many-body systems at scales smaller than the coherence length (to keep coherent and interference effects) but larger than the mean free path, or scattering length (to deal with diffusive behaviours). By providing information on fluctuating coherences across the systems, the quantum version of the macroscopic fluctuation theory is expected to code for properties -- beyond linear response transport (as the Landauer-Buttiker theory already do) -- dealing with interferences, entanglements, information spreading, etc., in such systems and which have obviously no classical counterpart. For classical systems, establishing the general formalism of the macroscopic fluctuation theory started by solving and understanding properties of simple model systems such as the exclusion processes. Our work aims at taking a similar route in the quantum case.

There are a few differences between our work and that of the reference [69]:
Ref.[69] deals with a model of dissipative spin chain experiencing random dephasing at each local site. Ref. [69] looks at the mean dynamics encoded into a Lindblad equation. In one hand, ref.[69] deals with system models more general than ours, in the sense that they are valid for any strength of the dephasing, but on the other hand it deals with less general questions in the sense that it only looks at the mean behaviour whereas the purpose of our work is to decipher the fluctuations present in this class of model systems. Our model provides a description of the dynamics at the stochastic level while the local dephasing in [69] focused on the mean Lindblad dynamics, i.e. the dynamics averaged over the noise. The integrability structure used in [69], via a mapping to a complex Hubbard model, only applies to the mean Lindblad dynamics. Even though it was shown by the authors and M. Bauer (ref. [65] in the text) that the stochastic version of the random dephasing model and the quantum SSEP (called the stochastic XY model in [65]) are connected in the limit of long time and strong dephasing, it is a priori not expected that the integrable structure deciphered in [69] could provide information beyond the mean properties.
However, there are strong incentives, some of which have been shown in this paper, that some notion of integrability should hold true for the quantum SSEP at the stochastic level. Decoding which integrability structure is behind the exact solution we described is yet an open problem (and we hope to come back to this problem). Once this last problem would have been solved, its solution combined with the result of [69] would suggest that the integrability structure inherent to the quantum SSEP admits an extension to the dephasing spin chain at the stochastic level (and not only at the mean Lindblad level). 

%

\end{document}